\documentclass[fleqn,10pt]{wlscirep}
\usepackage[utf8]{inputenc}
\usepackage[T1]{fontenc}
\usepackage{mathtools} 
\usepackage{bm}
\usepackage{dsfont,amsthm,amsbsy}
\usepackage{verbatim}
\usepackage{amssymb}
\usepackage{amsmath}
\usepackage{bbm}
\usepackage{graphicx}
\usepackage{epstopdf}
\usepackage{subfigure}
\usepackage{epsfig}
\usepackage{xcolor}
\usepackage{amsfonts}
\usepackage{mathrsfs}
\usepackage{sidecap}
\usepackage{lipsum}
\usepackage[toc,page,title,titletoc,header]{appendix}
\usepackage{hyperref}
\usepackage{resizegather}
\usepackage{tikz}
\usepackage{float}
\usepackage{mathbbol}
\usepackage[normalem]{ulem}
\usepackage{cancel}
\usepackage{upgreek}
\usepackage{wrapfig}

\newcommand{\bl}{\begin{aligned}}
\newcommand{\el}{\end{aligned}}

\def\be{\begin{equation}}
\def\ee{\end{equation}}

\def\bi{\begin{itemize}}
\def\ei{\end{itemize}}
\def\bn{\begin{enumerate}}
\def\en{\end{enumerate}}
\def\bea{\begin{eqnarray}}
\def\eea{\end{eqnarray}}
\def\no{\nonumber}
\def\ba{\begin{array}}
\def\ea{\end{array}}
\def\bd{\begin{displaymath}}
\def\ed{\end{displaymath}}

\renewcommand{\aa}[1]{{\color{blue} #1}}

\renewcommand{\aa}[1]{{\color{blue} #1}}

\usepackage{color}
\graphicspath{{.}{./Figures/}}
\begin{document}
%=====================================================
\title{Dynamics of decoherence in a noisy driven environment}
%=====================================================
\author[1,2,3*]{R. Jafari}
\author[4]{A. Asadian}
\author[5]{M. Abdi}
\author[6,7]{Alireza Akbari}

\affil[1]{Department of Physics, Institute for Advanced Studies in Basic Sciences (IASBS), Zanjan 45137-66731, Iran}
\affil[2]{School of Quantum Physics and Matter, Institute for Research in Fundamental Sciences (IPM), Tehran 19538-33511, Iran}
\affil[3]{Department of Physics, University of Gothenburg, SE 412 96 Gothenburg, Sweden}
\affil[4]{Department of Physics, Institute for Advanced Studies in Basic Sciences (IASBS), Zanjan 45137-66731, Iran}
\affil[5]{Wilczek Quantum Center, School of Physics and Astronomy, Shanghai Jiao Tong University, Shanghai 200240, China}
\affil[6]{Beijing Institute of Mathematical Sciences and Applications (BIMSA), Huairou District, Beijing 101408, P. R. China}
\affil[7]{Institut für Theoretische Physik III, Ruhr-Universität Bochum, 44801 Bochum, Germany}

\affil[*]{raadmehr.jafari@gmail.com}

\date{\today}

\begin{abstract}

We analyze the decoherence dynamics of a central spin coupled to a spin chain with a time-dependent noisy magnetic field, focusing on how noise influences the system's decoherence. 
Our results show that  decoherence due to the nonequilibrium critical dynamics of the environment is amplified in the presence of uncorrelated and correlated Gaussian noise. 
We demonstrate that decoherence factor consistently signals  the critical points, and exhibits exponential scaling with the system size, the square of noise intensity, and the noise correlation time at the critical points.
We find that strong coupling between the qubit and the environment leads to partial revivals of decoherence, which diminish with increasing noise intensity or decreasing noise correlation time. In contrast, weak coupling leads to monotonic enhanced decoherence.
The numerical results illustrate that, the revivals decay and scale exponentially with noise intensity. Moreover, the revivals increase and indicate linear or power law scaling with noise correlation time depends on how the correlated noise is fast or slow. 
Additionally, we explore the non-Markovianity of the dynamics, finding that it decays in the presence of noise but increases as the noise correlation time grows. 
%Our findings have potential applications in the noise spectroscopy of external signals.
%

\end{abstract}

\flushbottom
\maketitle
\thispagestyle{empty}
%-----------------------------------------------------------------------------
\section{Introduction}
Quantum correlations (QCs) are essential in quantum information science ~\cite{Barenco1,Pereira,Bera2018,Rao2013} and quantum computation~\cite{Barenco2,Grover,Jafari:2010aa,Mishra2018,Kaszlikowski2008}, as they are pivotal for understanding the inherent non-locality in quantum mechanics~\cite{Einstein, Bell,Sadhukhan2016}. During quantum information processing, quantum systems are inevitably affected by interactions with their surrounding environment. These interactions result in quantum decoherence, which is key to comprehend the transition from quantum to classical behavior~\cite{Zurek1,Kaiserbook,Jafari2017,Chanda2016,Jafari2017b,Jafari2015}.
In order to grasp the environment induced-decoherence, the model of a single central spin interacting with an environment, known as the central spin model (CSM)~\cite{Schliemann2002,Cucchietti2005,Korbicz2021,Cucchietti2006,Suzuki2016,Nag2012,Damski2011}, has been extended to the notion of quantum phase transition ~\cite{Quan2006,Jafari2017}. Within the CSM framework, an environmental quantum spin system (ESS) interacting with a central spin (CS) or qubit can be either 
time-independent or time-dependent.

In the time-independent case, the environmental spin system starts in its lowest energy state, whereas the central spin/qubit is initially in a pure state. The overall connection between the qubit and the environment is structured so that the ESS's initial ground state evolves along two distinct pathways, each governed by a different Hamiltonian. Although the qubit starts in a pure state, it has been shown that it loses almost all of its purity as the ESS approaches its quantum critical point~\cite{Quan2006,Jafari2017}.
Furthermore, in the context of time-dependent scenarios, as the ESS is gradually moved through its quantum critical point, there is a notable increase in decoherence. This amplification arises not only from the heightened susceptibility present near the critical point but also from the provoked excitations, suggesting an aspect that is dependent on the universality class. This phenomenon shows a remarkable similarity to the dynamics associated with defect formation seen in nonequilibrium phase transitions, as described by the Kibble-Zurek mechanism.~\cite{Kibble2007,zurek1985cosmological}.

Nevertheless, there has been a limited focus on the investigation of stochastically driven ESSs characterized by noisy Hamiltonians, and the influence of quantum coherence within these systems is still largely unexamined. 

{\color{black} Noise is a major obstacle in achieving goals in all quantum technologies, especially in the progression of quantum computing.\cite{Uhrig2007,Fink2013,Yuge2011}. 
In particular, by altering the outcome of a quantum dynamics through the disturbance that it introduces to the system parameters. Such disturbance can become significant which leads to information loss in qubits and a general deviation of the system from its intended state.
Due to the conceptual and technical complexities in dealing with the system plus environment fully quantum mechanically, an alternative approach is to simply consider that the effect of the environment is to introduce classical noise in the system’s degrees of freedom \cite{Budini2001,Yang2017,Montiel2013,Chenu2017,Spanner2009,Iwakura2017}. In other words, noise arises as an efficient way of describing the evolution of systems interacting with environments or external driven fields. Surmounting this challenge, is vital for advancements in quantum computing, with a goal to attain greater reliability and accuracy.}

In addition, in any real experiment, the simulation of the intended time-dependent Hamiltonian is inherently imperfect, and the presence of noisy fluctuations is inevitable. In essence, noise is pervasive and unavoidable in any physical system, exemplified by noise-induced heating that may result from amplitude variations in the lasers used to create the optical lattice ~\cite{Pichler,Chen2010,Zoller1981,Doria2011}. Consequently, it is essential to comprehend the effects of noise on Hamiltonian evolution to accurately forecast experimental results and to develop advanced configurations that are robust against noise effects ~\cite{Pichler,Marino2012,Marino2014,Dutta2016PRL,Bando2020,Chenu2017,Abdi2011}. 

The question addressed in this paper is: 
What are the effects of a noisy ESS on the decoherence of a central qubit when it is driven across the QCPs? 
Specifically, is there still a universal pattern in the dynamically induced decoherence, as measured by the decoherence factor (DF) of the CS?
{\color{black}More precisely, can the decoherencies due to proximity to a critical point and noise be distinguished?}

We demonstrate that decoherence factor decreases in the presence of both correlated (colored) and uncorrelated (white) Gaussian noises. 
This reduction exhibits exponential scaling with the system size, the square of  noise intensity, 
and the noise correlation time at the critical points. Moreover, in the case of strong environment-qubit coupling (large ramp time scale), decoherence exhibits revivals in both noiseless and noisy scenarios. These revivals scale exponentially with the square of noise intensity. However, in the presence of fast colored noise (small correlation time), the revivals scale linearly with the noise correlation time, whereas in slow noise (large correlation time), they exhibit a power-law scaling with noise correlation time. Additionally, we find that the measure of non-Markovianity decreases with noise intensity and increases linearly with the noise correlation time.

%____________________________________
%------------------------------------------------------------
\section{Theoretical model}\label{model}
%
%%%%%%%%%%%%%%%%%%%%%%%%%%%%%%%%%%%%%%%%%  Fig. 1  %%%%%%%%%%%%%%%%%%%%%%%%%%%%%%%%%%%%%%%%%%%
\begin{wrapfigure}{r}{0.5\textwidth}
\centering
\includegraphics[width=0.5\columnwidth, clip=true]{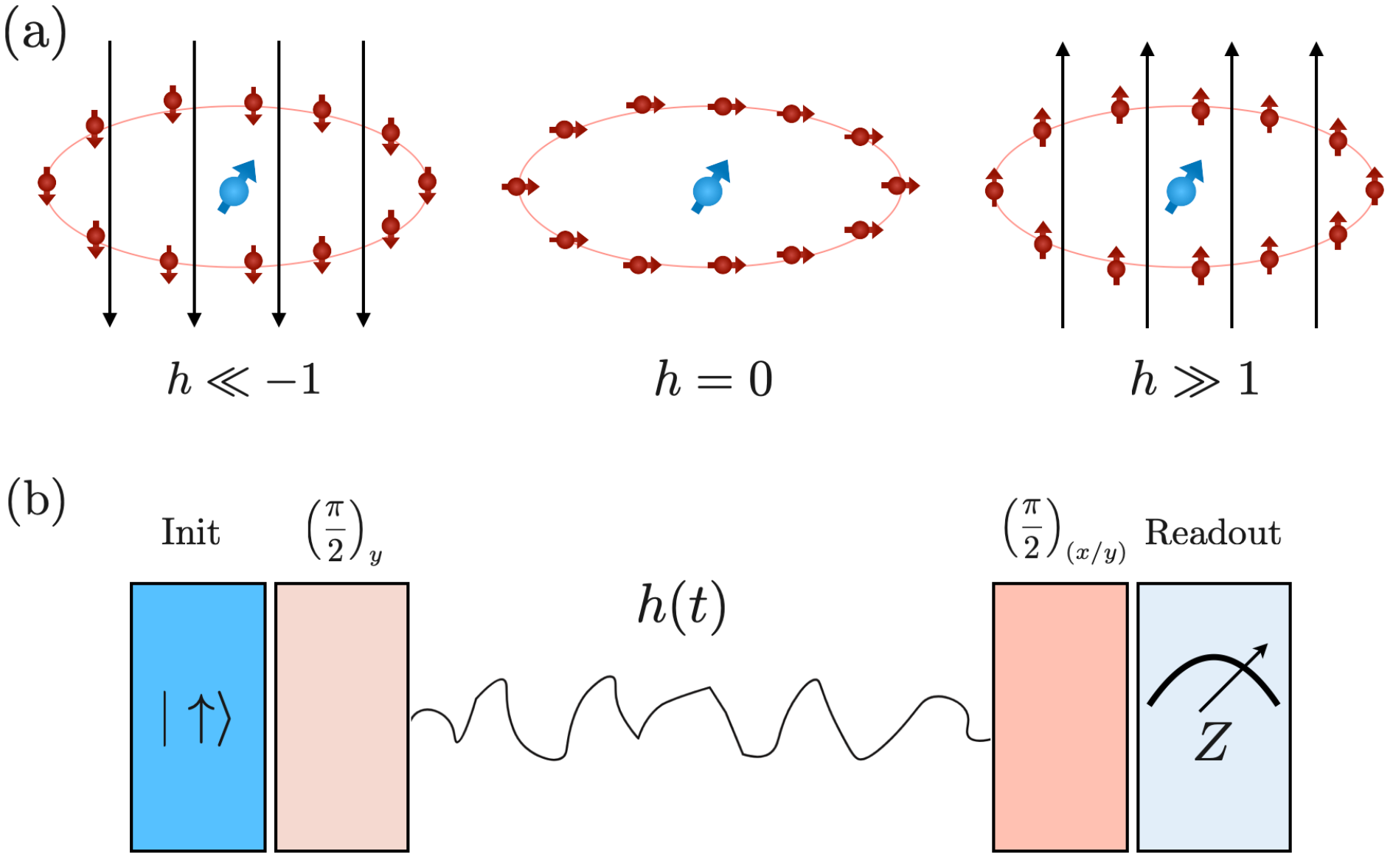}
\caption{
(a) The schematic diagram of spin qubit symmetrically coupled to the critical spin
environment described as a transverse field Ising chain. When the magnetic field is large 
($|h(t)| > 1$), (right and left panels), the environment is in a paramagnetic phase, 
where spins aligned with the field. In contrast, when the field is small ($|h(t)| < 1$;  (center panel), 
the spin chin enters a ferromagnetic phase, where spins ordered in $x$ or $-x$ direction.
(b) The decoherence factor can be directly measured through a Ramsey experiment, which involves a $\pi/2$ 
pulse sequence followed by a projective measurement in the $Z$-basis on the central qubit \cite{Bylander20,Zhang2020,Jurcevic2022,Asadian2014}.
}
\centering
\label{fig1}
\end{wrapfigure}
%%%%%%%%%%%%%%%%%%%%%%%%%%%%%%%%%%%%%%%%%%%%%%%%%%%%%%%%%%%%%%%%%%%%%%%%%%%%%%%%%%%%
%
%
The full Hamiltonian which considers a qubit coupled to a driven transverse field Ising chain (Fig.~\ref{fig1}), is expressed as \cite{Damski2011,Suzuki2016,Nag2012} 
\bea
\label{eq1}
\bl
{\cal H} = {\cal H}_E + {\cal H}_I + {\cal H}_q,
\el
\eea
where 
$$
{\cal H}_E = {\cal H}_E (h(t)) = 
-
\sum_{j=1}^N
 \Big(\sigma_j^{x}\sigma_{j+1}^{x} 
 +
  h(t)\sigma_j^{z}
  \Big),
$$
represents the time dependent transverse field Ising model in a ring configuration, 
%${\cal H}_I$ 
$$
{\cal H}_I ={\cal H}_I (\delta) = - \delta\sum_{j=1}^N \sigma_j^{z}\sigma^z_0,\
$$
describes the interaction between the surrounding spin ring and the central qubit, 
and 
%
%${\cal H}_q$ 
$$
{\cal H}_q = \sigma_0^z,
$$
corresponds to the Hamiltonian of the qubit.
Here, 
$\sigma^{\alpha = x, y, z}$ are Pauli matrices,
and 
$\delta$ represents the interaction strength between environment and the qubit. 
To search the effects of noise on the decoherence, we consider the noise added time dependent transverse field, i.e., $h(t)= h_0(t)+S(t)$, 
in which the noiseless transverse field $h_0(t)$ varying from an initial value $h_i$ at time $t_i$ to a final value $h_0(t)$ at time $t$
following the linear quench protocol $h_0(t) = t/\tau_Q$, where $\tau_Q$ is the ramp time scale, and $S(t)$ is the stochastic noise.
When the transverse field is time-independent and noiseless, $h_0(t) = h$, the ground state of the model is in the ferromagnetic phase for $|h| < 1$, 
while the system is in the paramagnetic phase for $|h| > 1$, with the phases separated by equilibrium quantum critical points at $h_c = \pm 1$~\cite{Pfeuty1970}.

Assuming the qubit is initially in a pure state at $h(t_i)$ \cite{Damski2011,Suzuki2016,Nag2012},
$$
|\phi(t_i)\rangle_{q}=c_u|\uparrow\rangle+c_d|\downarrow\rangle,
$$
% 
%
%%%%%%%%%%%%%%%%%%%%%%%  Fig. 2   %%%%%%%%%%%%%%%%%%%%%%%
\begin{figure*}
\begin{minipage}{\linewidth}
\centerline{\includegraphics[width=0.33\linewidth]{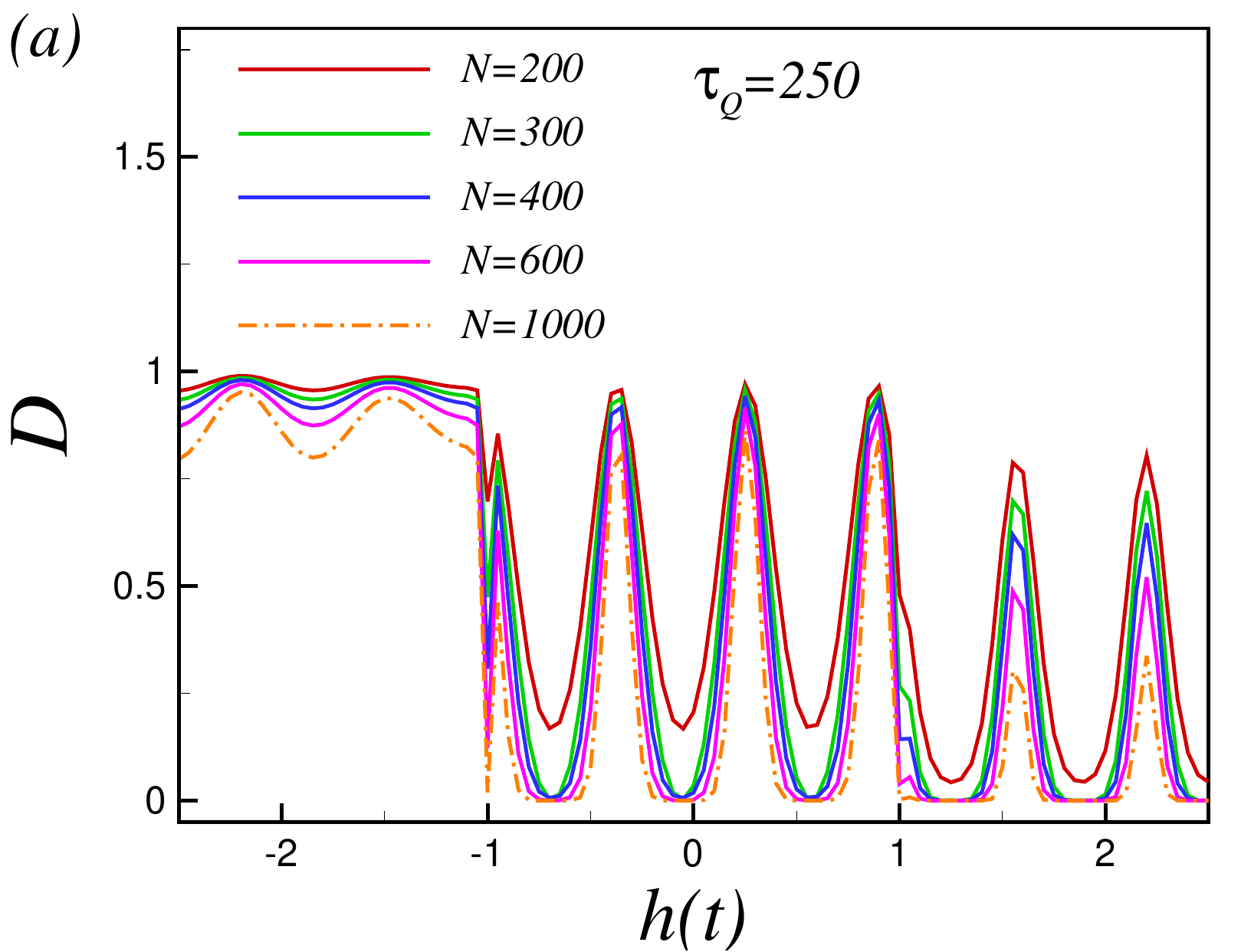}
\includegraphics[width=0.33\linewidth]{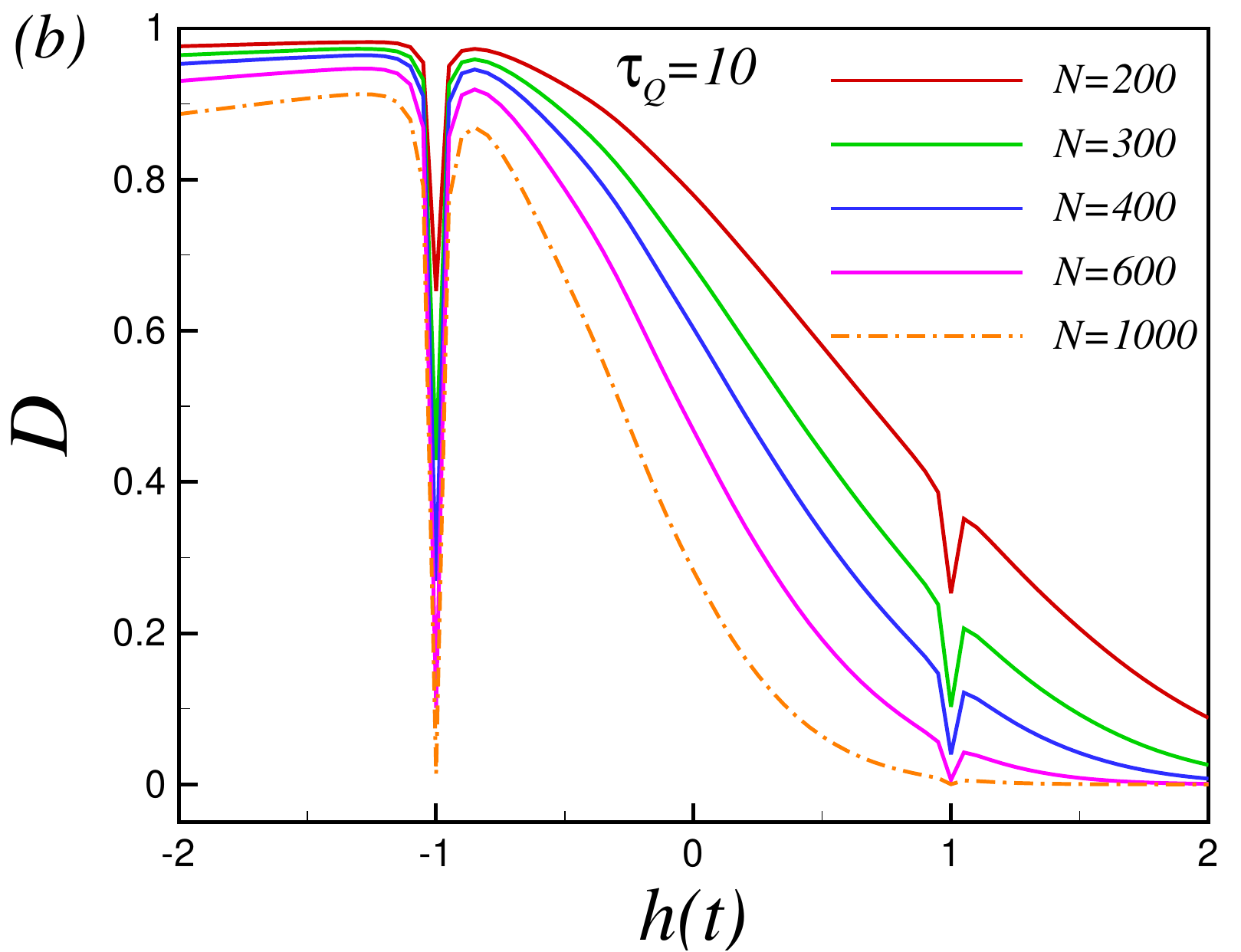}
\includegraphics[width=0.33\linewidth]{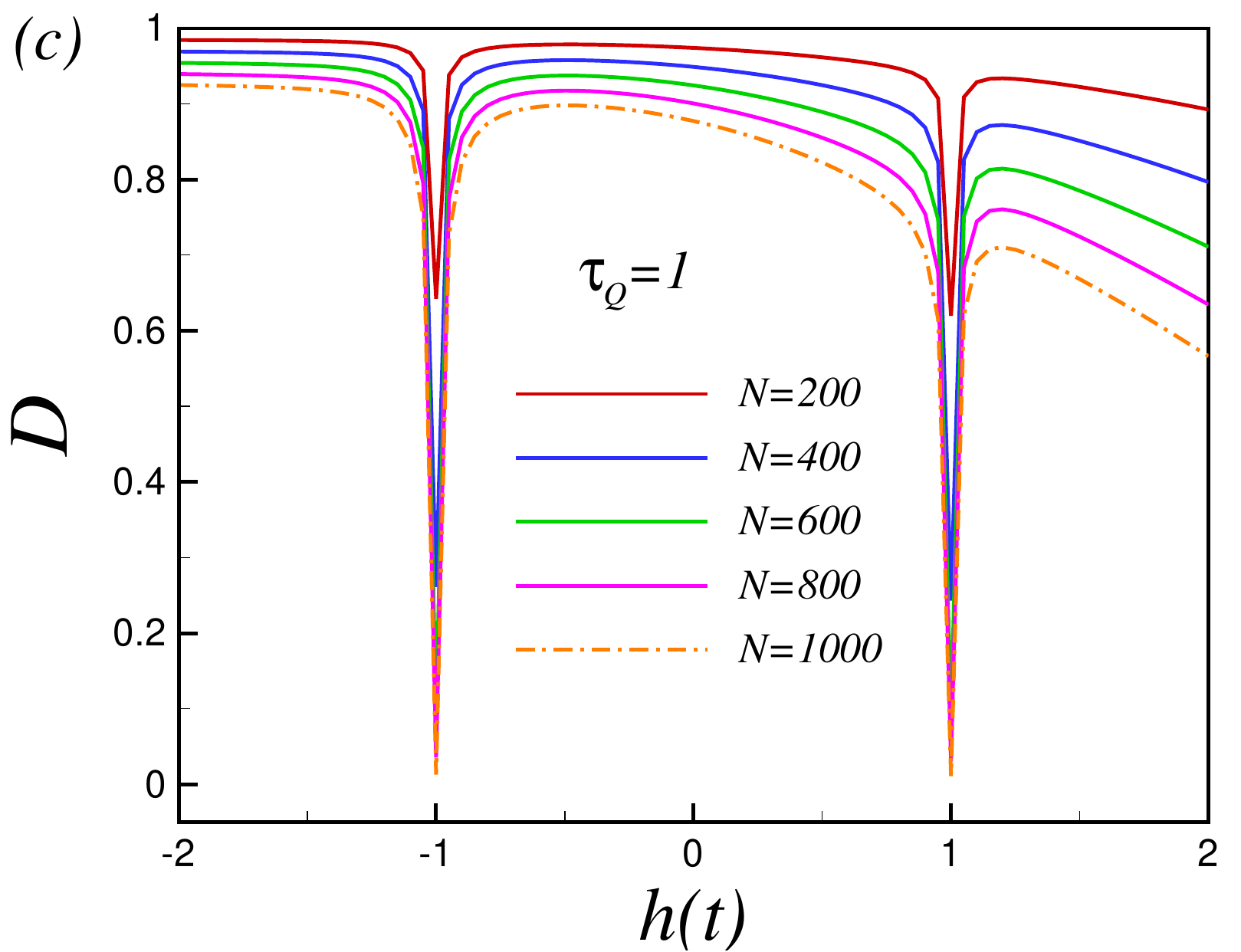}}
\centering
\end{minipage}
%===============================================================
\caption{ %(Color online) 
Noiseless decoherence, as measured by the visibility or decoherence 
function during a quench, is analyzed for $\delta = 0.01$ 
with varying ramp time scales $\tau_Q$ and system sizes $N$:
(a) For $\tau_Q = 250$, nearly perfect revivals of decoherence observed between two 
critical points. However, decoherence is reduced when the magnetic field 
swept through the second critical point, and wiped out for very large system sizes.
(b) For $\tau_Q = 10$, there is a monotonic decay of decoherence between the critical points. 
This behavior is observed when $\tau_Q$ exceeds the threshold value of $\pi/(16\delta)$, 
indicating that $\tau_Q$ is approximately around this threshold.
(c) The decoherence behavior during the quench for $\tau_Q = 1$ is illustrated, 
highlighting the differences in decoherence loss compared to the other time scales.
}
\label{figAPA1}
\end{figure*}
%%%%%%%%%%%%%%%%%%%%%%%%%%%%%%%%%%%%%%%%%%%%%%%%%%%%%%%
%  
with coefficients satisfying $|c_u|^2 + |c_d|^2 = 1$, and the 
environment, $E$, is in the ground state denoted by 
$ |\varphi(t_i)\rangle_E$,  the total wave function of the composite system at time $t_i$ can then be written in the direct product form \cite{Damski2011,Suzuki2016,Nag2012}
%
%%%%%%%%%%%%%%%%%%%%%%%%%%%%%%%% Eq. 3 %%%%%%%%%%%%%%%%%%%%%%%%%%%%%%%%
\begin{equation}
\label{eq3}
|\Psi(t_i)\rangle = |\phi(t_i)\rangle_q \otimes |\varphi(t_i)\rangle_E      
.
\end{equation}
%%%%%%%%%%%%%%%%%%%%%%%%%%%%%%%%%%%%%%%%%%%%%%%%%%%%%%%%%%%%%%%%%%%%
%
%
It is straightforward to show that, the total wave function at an instant $t$ is given by~\cite{Damski2011,Suzuki2016,Nag2012}
%
%%%%%%%%%%%%%%%%%%%%%%%%%%%%%%%% Eq. 4 %%%%%%%%%%%%%%%%%%%%%%%%%%%%%%%%
\bea
\label{eq4}
|\Psi(t)\rangle = c_d|\downarrow\rangle \otimes |\varphi^{-}(t)\rangle + c_u|\uparrow\rangle \otimes |\varphi^{+}(t)\rangle,
\eea
%%%%%%%%%%%%%%%%%%%%%%%%%%%%%%%%%%%%%%%%%%%%%%%%%%%%%%%%%%%%%%%%%%%%
%
where
\begin{equation}
|\varphi^{\pm}(t)\rangle = \hat{T} \exp\left[-i \int_{t_i}^t
 {\cal H}_E
 \Big(
 h(t) \pm \delta
 \Big)
  \, dt\right] |\varphi^{\pm}(t_i)\rangle,
\end{equation}
and $\hat{T}$ denotes the time-ordering operator. 
Since  interaction Hamiltonian commutes with the central spin Hamiltonian, the basis $\{|\uparrow\rangle, |\downarrow\rangle\}$ are stationary states. 
Thus, the evolution of the entire system simplifies to the dynamics of two Ising branches, each evolving in an effective magnetic field given by 
$h_{\rm eff}(t)=h(t) \pm \delta$.
Consequently, the spin-dependent evolution of the environmental states is governed by \cite{Damski2011,Suzuki2016,Nag2012}
%
%%%%%%%%%%%%%%%%%%%%%%%%%%%%%%%% Eq. 5 %%%%%%%%%%%%%%%%%%%%%%%%%%%%%%%%
\begin{equation}
i \frac{\partial}{\partial t} \left|\varphi^\pm(t)\right\rangle =
 {\cal H}_E
 \Big(
 h(t) \pm \delta
 \Big)
  \left|\varphi^\pm(t)\right\rangle.
\label{eq5}
\end{equation}
%%%%%%%%%%%%%%%%%%%%%%%%%%%%%%%%%%%%%%%%%%%%%%%%%%%%%%%%%%%%%%%%%%%%
%
%\aa{In the following section we present the state evolution by reviewing the solution technique to the Ising model.}

Given the qubit in the ${|\uparrow\rangle, |\downarrow\rangle}$ basis, the reduced density matrix is described by~\cite{Damski2011,Suzuki2016,Nag2012}
%
%%%%%%%%%%%%%%%%%%%%%%%%%%%%%%%% Eq.  %%%%%%%%%%%%%%%%%%%%%%%%%%%%%%%%
{\small
\begin{equation}
\bl
\rho_q(t)= {\rm Tr}_E
\Big[
|\psi(t)\rangle\langle\psi(t)|
\Big]
=
\left(
\begin{array}{cc}
|c_u|^{2} &   c_u c_d^* d^*(t) \\
c_u^* c_d d(t) & |c_d|^{2}  
\end{array}
\right),
\el
\end{equation}
}
%%%%%%%%%%%%%%%%%%%%%%%%%%%%%%%%%%%%%%%%%%%%%%%%%%%%%%%%%%%%%%%%%%%%
%
where $d(t) = \langle\varphi^+(t)|\varphi^-(t)\rangle$ captures the coherency of the spin. We focus on studying its squared modulus, $|d(t)|^2$, which is known as the decoherence factor, to analyze the time evolution of decoherence \cite{Damski2011,Suzuki2016,Nag2012}:
\bea
\label{eq7} 
D = |d(t)|^2 = |\langle\varphi^+(t)|\varphi^-(t)\rangle|^2.
\eea
Not to be confused with the term noise, it is preferable to refer to this property as visibility, which measures the interference contrast.
It is worth noting that the entanglement entropy, concurrence, and maximum quantum Fisher information are directly connected to the decoherence 
factor~\cite{Sun:2007aa,Sun:2010aa,Costi:2003aa}.
%
%The decoherence factor also quantifies the entanglement between the spin and the chain, where
The qubit is fully decohered when $D = 0$, but it stays in a pure superposition state when $D = 1$.
In the unique situation where $c_u = c_d = 1/\sqrt{2}$, the density matrix assumes the subsequent straightforward form
%
%%%%%%%%%%%%%%%%%%%%%%%%%%%%%%%%%%% Eq. 8 %%%%%%%%%%%%%%%%%%%%%%%%%%%%%%%%%%%
{\small
\begin{equation}
\bl
\rho_q(t) =
\frac{1}{2}\left(
\begin{array}{cc}
1 & d^*(t) \\
d(t) & 1  
\end{array}
\right)
= \frac{1}{2}\Big(\mathbbm{1} + \langle\sigma_0^x\rangle\sigma_0^x + \langle\sigma_0^y\rangle\sigma_0^y\Big).
\el
\end{equation}
}
%%%%%%%%%%%%%%%%%%%%%%%%%%%%%%%%%%%%%%%%%%%%%%%%%%%%%%%%%%%%%%%%%%%%%%%%%%%%%
%
%
From this, it becomes evident that one can directly measure $D$ by evaluating the Pauli $x$ and $y$ observables of the central spin. Specifically, $d(t) = \langle \sigma_0^x \rangle + i \langle \sigma_0^y \rangle$. The standard experimental scheme to perform this measurement is known as the Ramsey measurement, which involves suitable pulse sequences followed by a projective measurement in the $Z$-basis \cite{Bylander20,Zhang2020,Jurcevic2022,Asadian2014}, as illustrated in Fig.~\ref{fig1}(b). 
In this scheme, the first pulse performs a $\pi/2$-rotation of the initial qubit state around the $y$-axis. The second pulse applies another $\pi/2$-rotation, but around either the $x$-axis or the $y$-axis, depending on whether $\sigma_0^x$ or $\sigma_0^y$ is to be measured. The outcome statistics are then read out in the $Z$-basis to determine the expectation values of $\sigma_0^y$ or $\sigma_0^x$.

%
%%%%%%%%%%%%%%%%%%%%%%%  Fig.3   %%%%%%%%%%%%%%%%%%%%%%%
\begin{figure*}
\begin{minipage}{\linewidth}
\centerline{\includegraphics[width=0.33\linewidth]{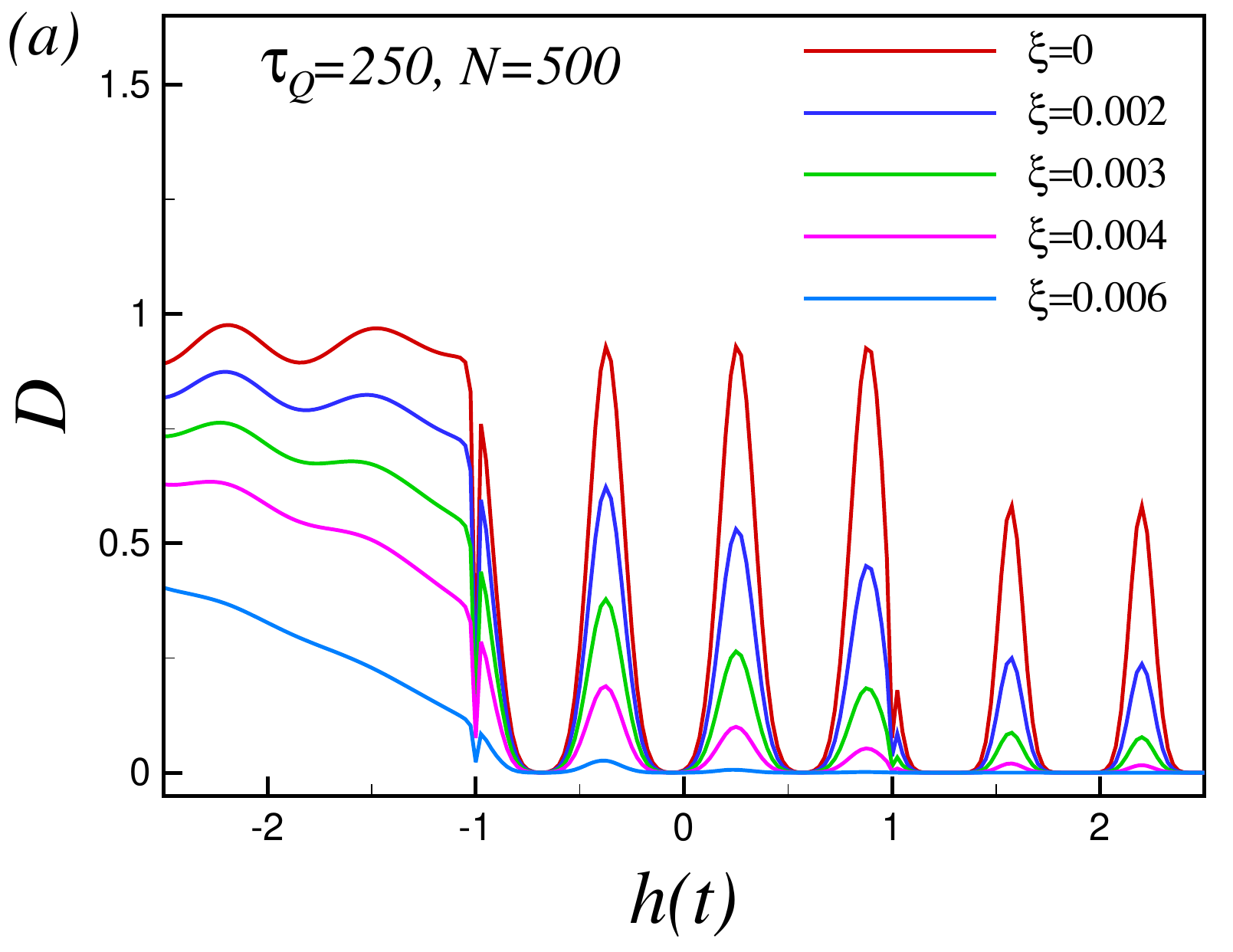}
\includegraphics[width=0.33\linewidth]{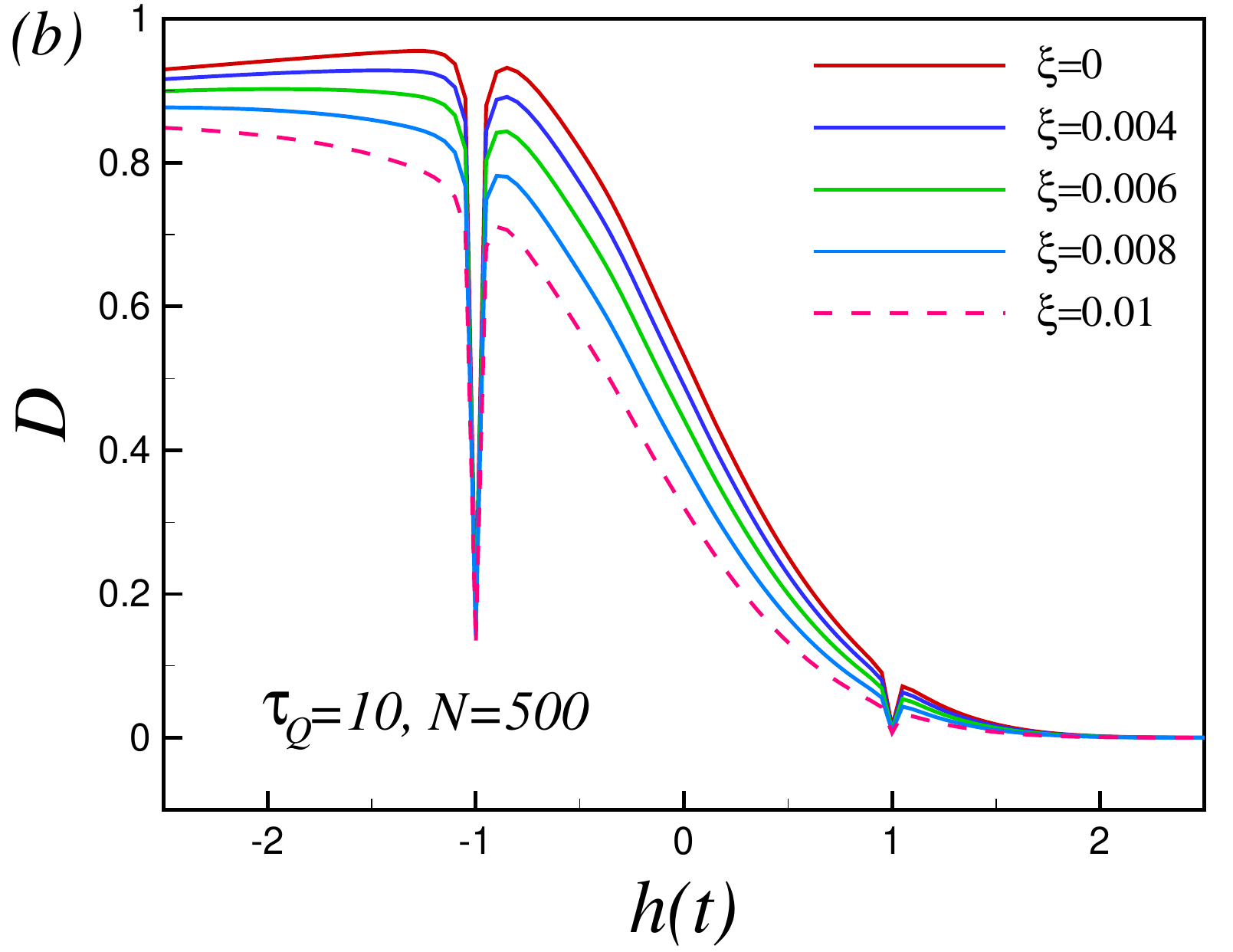}
\includegraphics[width=0.33\linewidth]{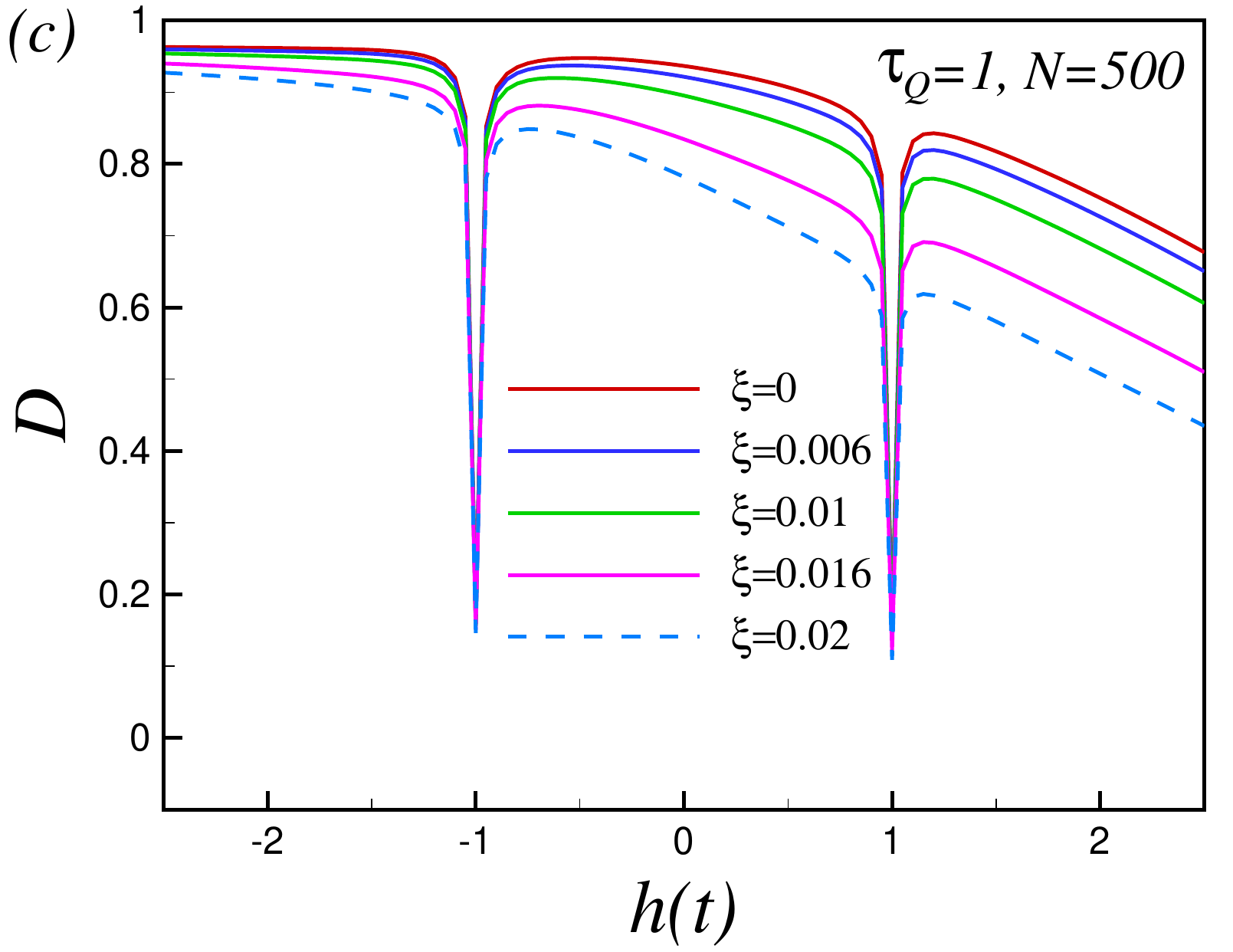}}
\centering
\end{minipage}
%===============================================================
\caption{
%(Color online) 
The noisy decoherence during a ramp up the magnetic field in the presence of
the white noise, for $\delta=0.01$, $N=500$ and different values of $\tau_Q$ and 
noise intensity: (a) As seen, revivals of decoherence which appear 
between two critical points for the ramp time scale $\tau_Q=250$, are still partially 
present in the presence of the white noise. 
(b) Monotonic decay of decoherence between the critical points for $\tau_Q=10$ 
increases in the presence of the white noise. 
(c) Illustrates decay of the noisy decoherence during the quench for $\tau_Q=1$.}
\label{fig2}
\end{figure*}
%%%%%%%%%%%%%%%%%%%%%%%%%%%%%%%%%%%%%%%%%%%%%%%%%%%%%%%
%  

%%%%%%%%%%%%%%%%%%%%%%%%%%%%%%%%%%%%%%%%%%%%%%%%%%%%%%%%%%%%%%%%%%%%%%%%%%%%%%%%%%%%%%%%%%%%%%%%%%%

\subsection{Decoherence factor and time dependent Schrödinger equation}

The behavior of the decoherence factor (visibility) can be determined by solving the time-dependent Schrödinger 
equation (Eq. \ref{eq5}). 
Through the application of Jordan-Wigner fermionization and use of Fourier transformation ~\cite{lieb_two_1961,Jafari2012}, 
the Hamiltonian presented in Eq. (\ref{eq1}) can be reformulated as the sum of $N/2$ non-interacting terms \cite{Damski2011,Suzuki2016,Nag2012}:

%
%%%%%%%%%%%%%%%%%%%%%%%%%%%%%%%% Eq. %%%%%%%%%%%%%%%%%%%%%%%%%%%%%%%%%%%%%%%%%%
\bea
\label{eq:Hk}
H(t) = \sum_{k} H_{k}(t),
\eea
%%%%%%%%%%%%%%%%%%%%%%%%%%%%%%%%%%%%%%%%%%%%%%%%%%%%%%%%%%%%%%%%%%%%%%%%%%%%%%
%
with
%
%%%%%%%%%%%%%%%%%%%%%%%%%%%%%%%%%%%%%%%%%  Equation.6  %%%%%%%%%%%%%%%%%%%%%%%%%%%%%%%%%%%%%%%%%%%
\bea
\label{eq6}
\bl
H_{k}(t)=
[-(h(t)\pm\delta-\cos k)](c_{k}^{\dagger}c_{k}+c_{-k}^{\dagger}c_{-k}) +\sin(k)(c_{k}^{\dagger}c_{-k}^{\dagger}+c_{k}c_{-k}),
\el
\eea
%%%%%%%%%%%%%%%%%%%%%%%%%%%%%%%%%%%%%%%%%%%%%%%%%%%%%%%%%%%%%%%%%%%%%%%%%%%%%%%%%%%%
%
where $c_{k},~ (c_{k}^{\dagger})$ represents spinless fermionic annihilation  (creation) operator, 
and the wave number $k$ is given by $k = (2m - 1)\pi / N$, with $m$ ranging from $1$ to $N/2$, and $N$ denotes 
the total number of spins (or sites) in the chain {\color{black}(mathematical details of analysis can be found in Supplemental Material)}.
The Bloch single-particle Hamiltonian $H_{k}(t)$ can be expressed as \cite{Damski2011,Suzuki2016,Nag2012}
%
%%%%%%%%%%%%%%%%%%%%%%%%%%%%%%%%%%%%%%%%%  Eq. 7  %%%%%%%%%%%%%%%%%%%%%%%%%%%%%%%%%%%%%%%%%%%
\bea
\label{eq7}
H_{k}(t) =
\left(
\begin{array}{cc}
h_{k}(t) & \Delta_{k} \\
\Delta_{k} & -h_{k}(t) \\
\end{array}
\right),
\eea
%%%%%%%%%%%%%%%%%%%%%%%%%%%%%%%%%%%%%%%%%%%%%%%%%%%%%%%%%%%%%%%%%%%%%%%%%%%%%%%%%%%%
%
where $\Delta_{k} = \sin(k)$ and $h_{k}(t) = -( h(t) \pm\delta - \cos(k))$.
Thus, the time-dependent Schrödinger equation (Eq. (\ref{eq5})) {\color{black}with $|\varphi_{k}^{\pm}(t)\rangle=(v^{\pm}_{k}, u^{\pm}_{k})^{T}$} can be expressed as
%
%%%%%%%%%%%%%%%%%%%%%%%%%%%%%%%% Eq. 8 %%%%%%%%%%%%%%%%%%%%%%%%%%%%%%%%
\begin{eqnarray}
\label{eq8}
\bl
i \frac{d}{dt} v^{\pm}_{k} =
& -(h(t) \pm \delta - \cos k) v^{\pm}_{k} + \sin k \, u^{\pm}_{k},
 \\
i \frac{d}{dt} u^{\pm}_{k} =
& (h(t) \pm \delta - \cos k) u^{\pm}_{k} + \sin k \, v^{\pm}_{k}.
\el
\end{eqnarray}
%%%%%%%%%%%%%%%%%%%%%%%%%%%%%%%%%%%%%%%%%%%%%%%%%%%%%%%%%%%%%%%%%%%%
%
%
Within this framework, one can derive that
%
%%%%%%%%%%%%%%%%%%%%%%%%%%%%%%%% Eq. 9 %%%%%%%%%%%%%%%%%%%%%%%%%%%%%%%%
\begin{equation}
D(t) = \prod_{k>0} F_k(t);
\quad 
 F_k(t) = \left| u_{k}^{+*}(t) u_{k}^-(t) + v_{k}^{+*}(t) v_{k}^-(t) \right|^2,
\label{eq9}
\end{equation}
%%%%%%%%%%%%%%%%%%%%%%%%%%%%%%%%%%%%%%%%%%%%%%%%%%%%%%%%%%%%%%%%%%%%
%
where $F_k(t)$ captures the dynamics of decoherence from the perspective of momentum space~\cite{Damski2011,Suzuki2016,Nag2012}. 

%
%%%%%%%%%%%%%%%%%%%%%%%  Fig.4   %%%%%%%%%%%%%%%%%%%%%%%
\begin{figure*}
\begin{minipage}{\linewidth}
\centerline{\includegraphics[width=0.33\linewidth]{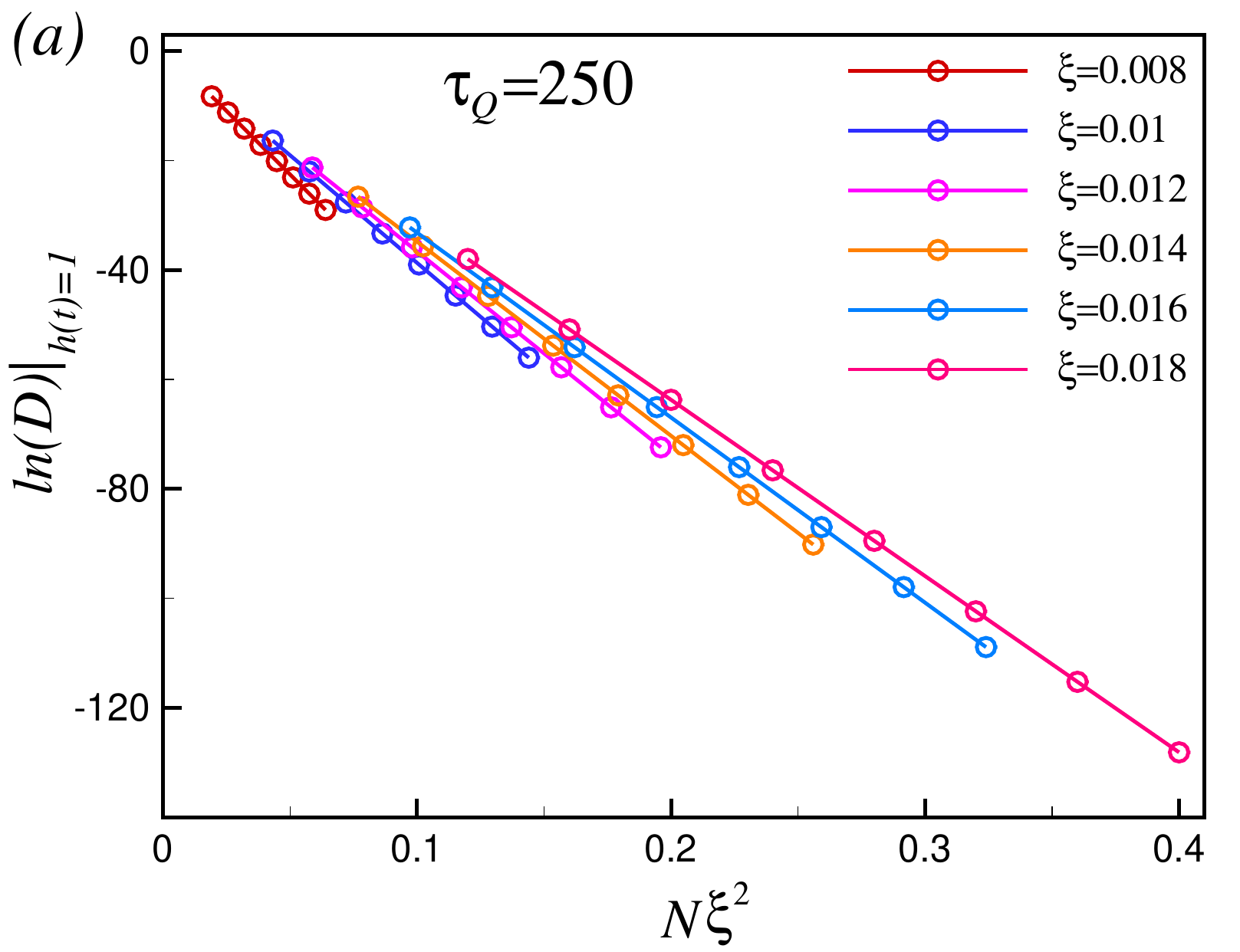}
\includegraphics[width=0.33\linewidth]{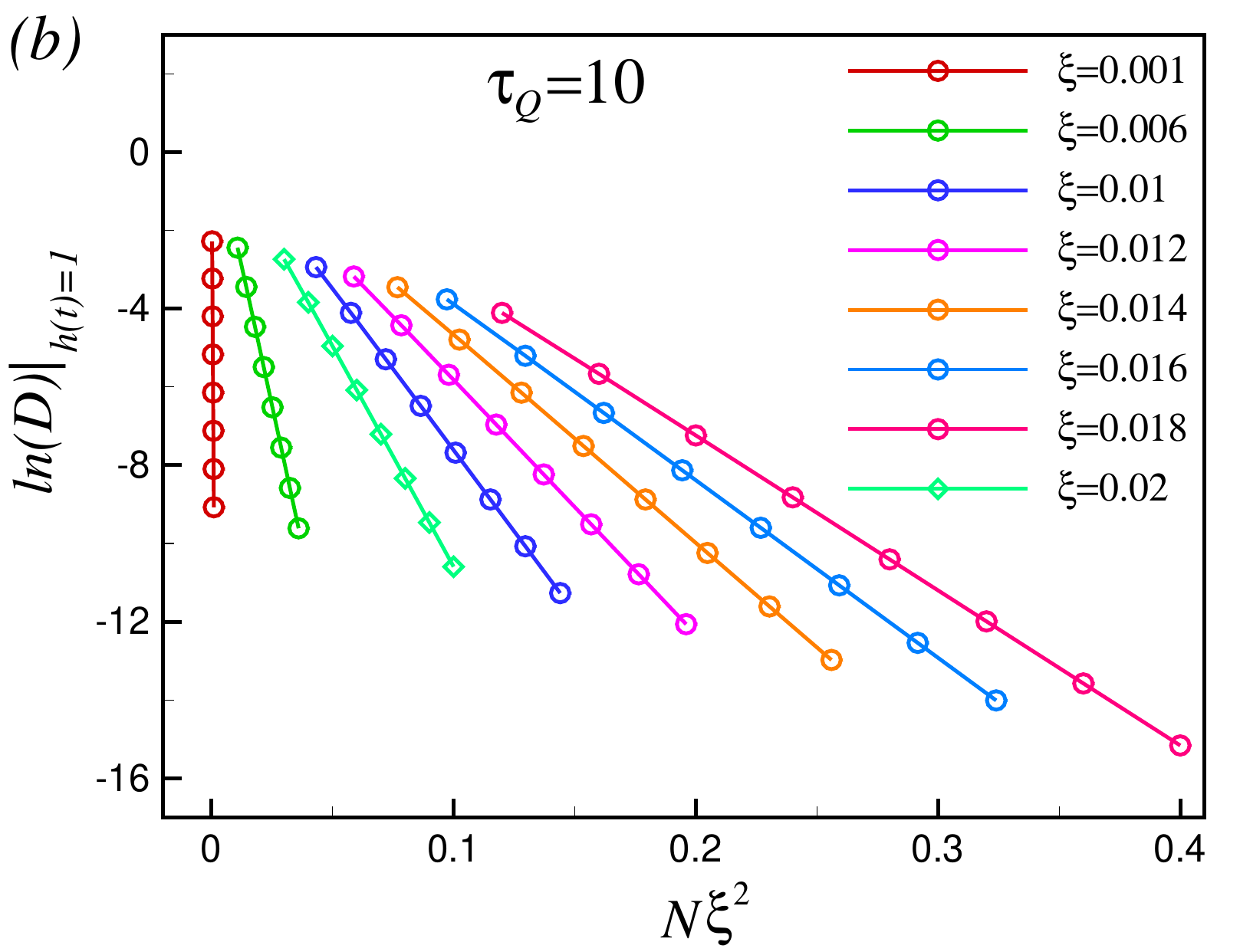}
\includegraphics[width=0.33\linewidth]{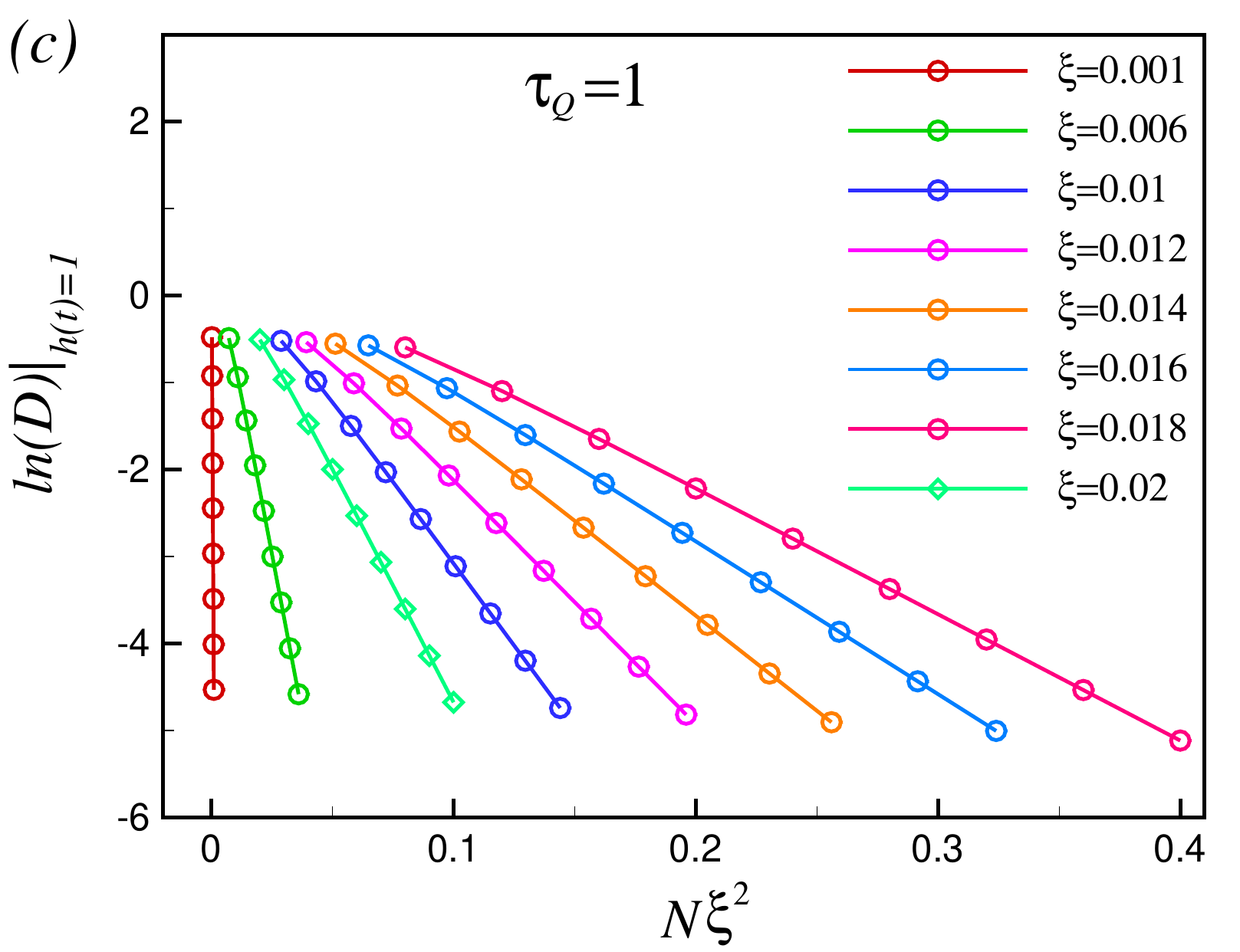}}
\centering
\end{minipage}
%===============================================================
\caption{
%(Color online)
Scaling of the minimum decoherence at the critical point $h_c=1$ versus 
$N\xi^2$ for the  different ramp time scales: 
(a) $\tau_Q=250$,
(b) $\tau_Q=10$, 
(c) $\tau_Q=1$.}
\label{fig3}
\end{figure*}
%%%%%%%%%%%%%%%%%%%%%%%%%%%%%%%%%%%%%%%%%%%%%%%%%%%%%%%
%

It is worthy to note that, in the absence of noise $S(t)=0$, it can be demonstrated that the coupled deferential equations in Eq. (\ref{eq8}) are exactly solvable ~\cite{Damski2005,Damski2011,Jafari2024a,Zamani2024,Baghran2024,Suzuki2016,Nag2012}. While in the presence of noise, the ensemble average of $u_{k}^{\pm}(t)$ and $v_{k}^{\pm}(t)$ 
can be calculated numerically using the master equation ~\cite{Budini2001,Kiely2021,Jafari2024a,Zamani2024,Baghran2024,Sadeghizade2025}. In the following we first review the dynamics of decoherence factor in the noiseless case and then we search the effects of noise on the dynamics of decoherence factor.  

%%%%%%%%%%%%%%%%%%%%%%%%%%%%%%%%%%%%%%%%%%%%%%%%%%%%%%%%%%%%%%%%%%%%%%%%%%%%%%%%%%%%%%%%%%%%%%%%%%%%%%%%%%%%%%%%%%%%%%%%%%%%%%%%%%%%%%%%

\section{Noiseless decoherence factor}
To study the dynamics of the decoherence, we initially prepare the system in its ground state at $t_i\rightarrow-\infty$ ($h_0(t) \ll -1$), 
and then ramp up the magnetic field in such a way that the system crosses the ferromagnetic phase ($-1 < h_0(t) < 1$), to enter the other paramagnetic phase ($h_0(t) > 1$).
In such a case the quench field crosses the critical points where located at $h_c = \pm 1$.

As the transverse field is ramped up, the responsive of environment to external field increases. 
This heightened sensitivity leads to enhanced decoherence, which is observed as a gradual reduction in the decoherence factor 
$D$, beyond the critical point; see Fig.~\ref{figAPA1}. In the noiseless case, it has been shown that the decoherence factor 
in the paramagnetic phase is approximately given by~\cite{Damski2011,Suzuki2016,Nag2012}
%
%%%%%%%%%%%%%%%%%%%%%%%%%%%%%%%% Eq. 10 %%%%%%%%%%%%%%%%%%%%%%%%%%%%%%%%
\begin{equation}
D(t)\approx\exp\left(-\frac{N\delta^2}{4h(t)^2(h(t)^2-1)}\right).
\label{eq10}
\end{equation}
%%%%%%%%%%%%%%%%%%%%%%%%%%%%%%%%%%%%%%%%%%%%%%%%%%%%%%%%%%%%%%%%%%%%
%
As the transverse field crosses the first critical point, $h(t)>-1$, the adiabatic evolution breaks down which leading to acceleration of decoherence; 
appears as a substantial reduction in decoherence factor.
As a result, the dynamics of decoherence influenced by the two effects: (i) the excitations in the environments due to the crossing the critical point,  
(ii) the perturbation which amplified by enhancement of environment's sensitivity close to the quantum critical points \cite{Zamani2024,Jafari2024a,Baghran2024}.  

As the driven transverse field crosses the first critical point at $h_c = -1$, decoherence either partially revives [Fig.~\ref{figAPA1}(a)] or decays monotonically [Figs.~\ref{figAPA1}(b)~and~\ref{figAPA1}(c)] in the region between two critical points. These behaviors arise from the non-adiabatic dynamics of the system near the critical point, as the band gap closes for $k = \pi$ at $h_c = -1$, where large $k$ modes are excited. The large $k$ modes, particularly those with $k \sim \pi - \hat{k}$ with 
$ \hat{k} \sim \tau_Q^{-1/2}$ experience significant excitation ~\cite{Damski2011,Suzuki2016,Nag2012}, while the small $k$ modes, for which $k \ll \pi - \hat{k}$, evolve adiabatically through the critical point.
Additionally, partial revivals of decoherence between the critical points have been observed when the environment-qubit coupling is sufficiently strong, specifically when $\delta \gg \pi/(16\tau_Q)$ ~\cite{Damski2011,Suzuki2016,Nag2012}; see Fig.~\ref{figAPA1}(a). These revivals manifest within the magnetic field, $h(t)$, domain with a period of $\pi/(4\tau_Q\delta)$~\cite{Damski2011,Suzuki2016,Nag2012}. In contrast, in the weak coupling regime where $\delta \ll \pi/(16\tau_Q)$, the decoherence factor exhibits a monotonic decrease, as illustrated in Fig.~\ref{figAPA1}(b).
Moreover, more complex decoherence dynamics arise when the driven field inverses polarity and takes the system through the second critical point at $h_c = 1$. 
At this critical point, where the gap closing occurs at $k = 0$ the modes with $k \sim \hat{k}$ become excited. 

In the next section we will study the effect of noise on the dynamics of the decoherence factor using the numerical
solution of the exact master equation. 
%
%%%%%%%%%%%%%%%%%%%%%%%  Fig.5   %%%%%%%%%%%%%%%%%%%%%%%
\begin{figure*}
\begin{minipage}{\linewidth}
\centerline{\includegraphics[width=0.33\linewidth]{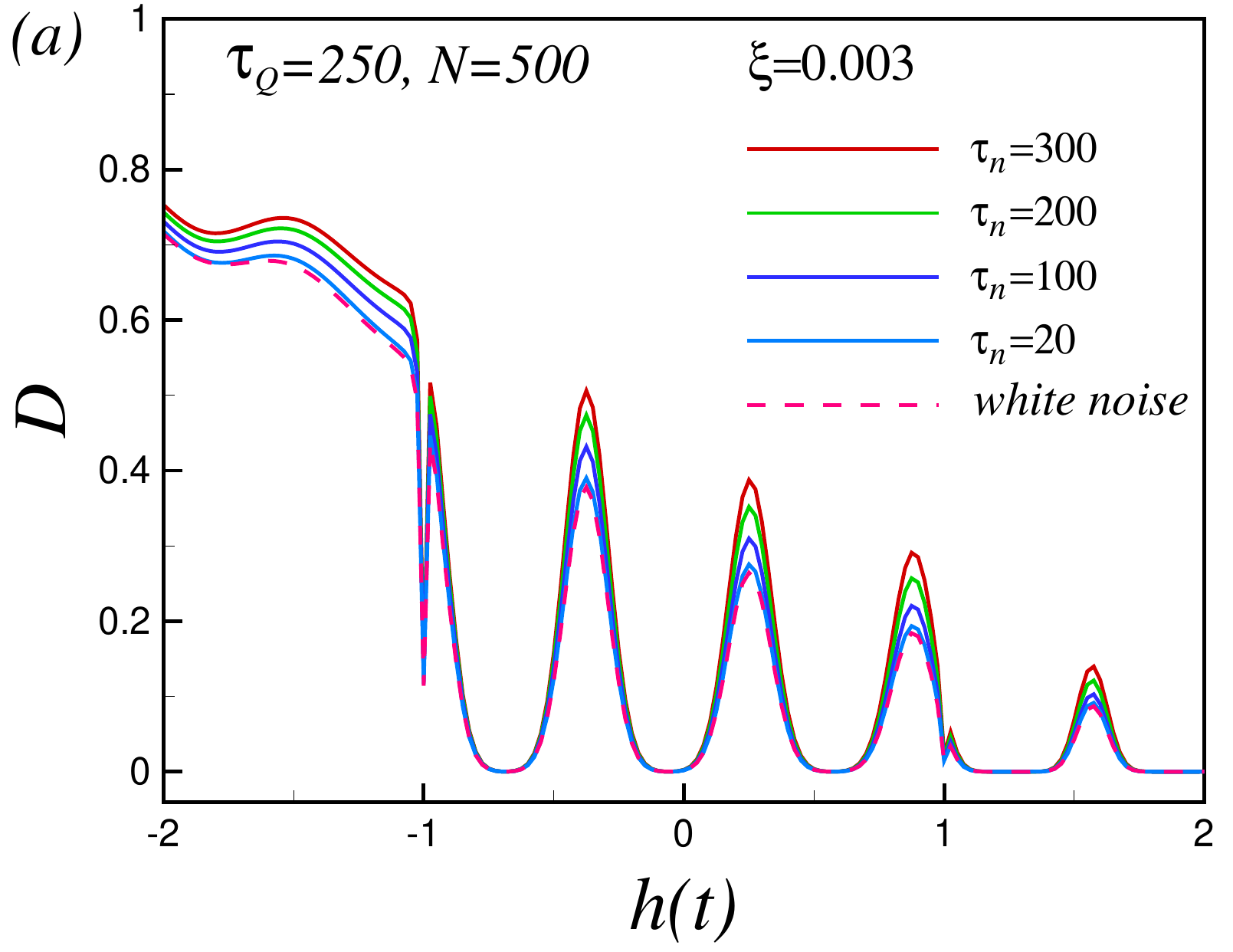}
\includegraphics[width=0.33\linewidth]{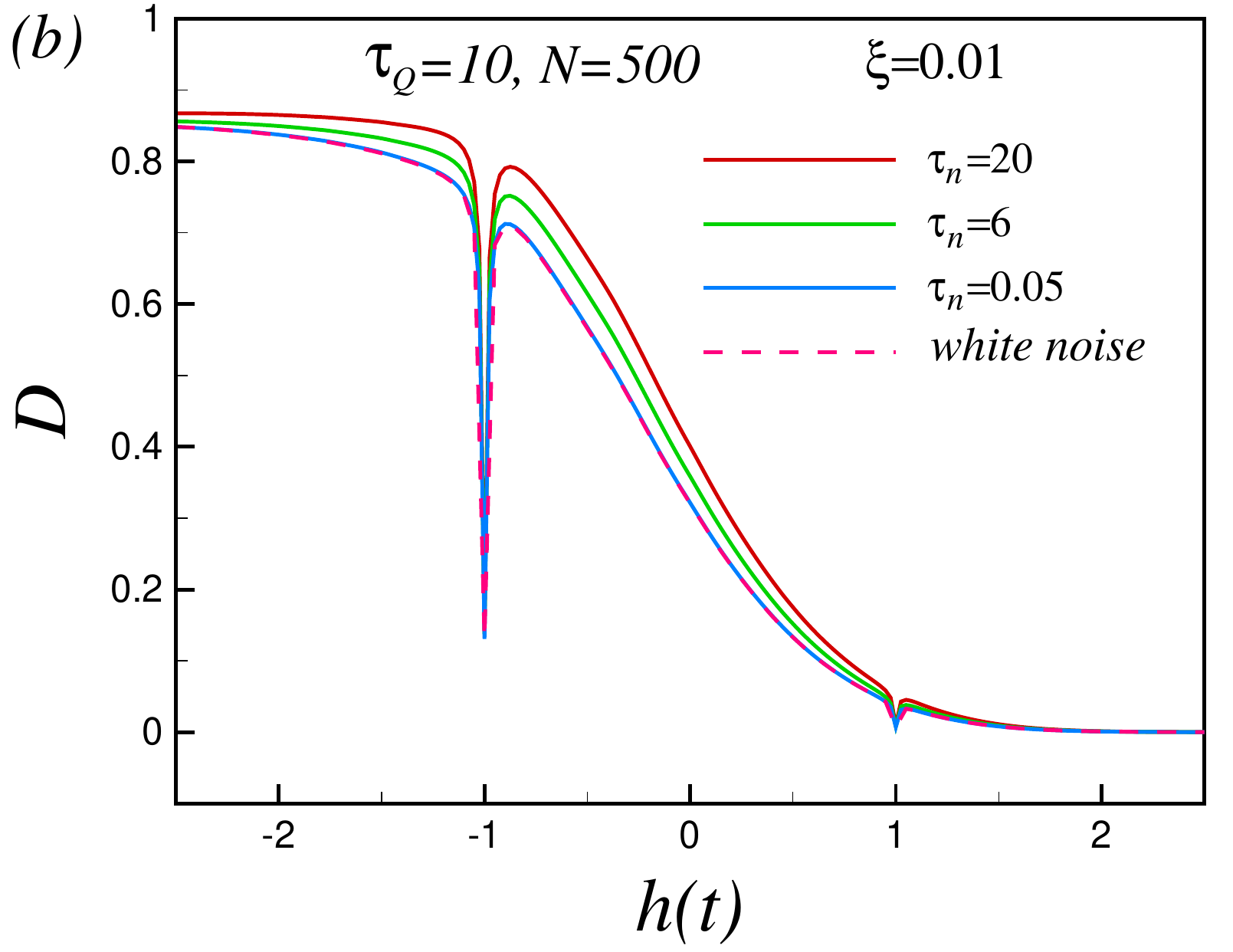}
\includegraphics[width=0.33\linewidth]{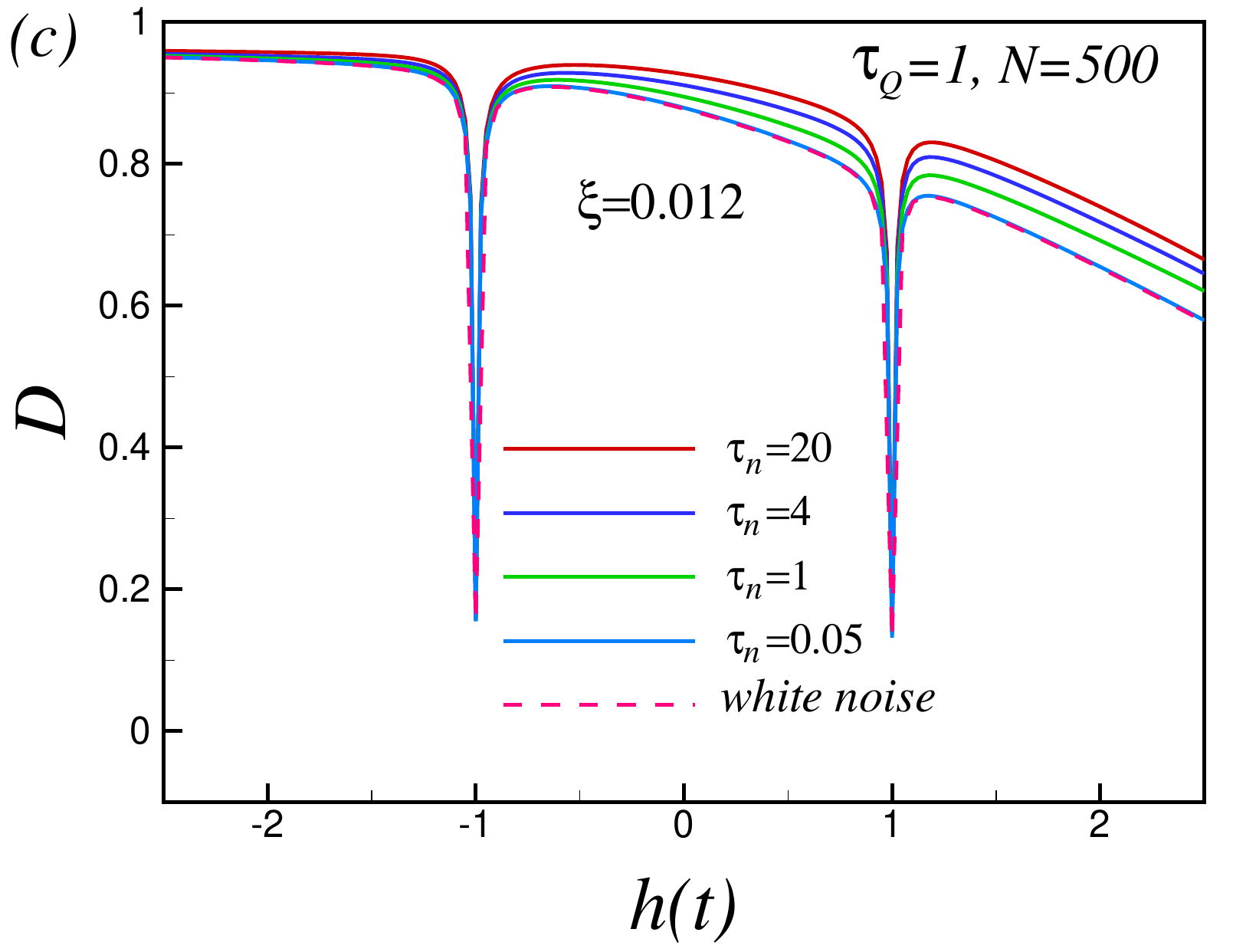}}
\centering
\end{minipage}
%===============================================================
\caption{
%(Color online)
The noisy decoherence during a quench in the presence of colored noise, with parameters $\delta=0.01$ and $N=500$:
(a) For $\tau_Q=250$ and $\xi=0.003$, revivals of decoherence observed between critical points in the ramp time scale $\tau_Q=250$ are also partially visible in the presence of colored noise.
(b) For $\tau_Q=10$ and $\xi=0.01$, there is a monotonic decay of decoherence between critical points, which becomes more pronounced with an increase in $\tau_n$.
(c) For $\tau_Q=1$ and $\xi=0.012$, the decay of noisy decoherence is shown for a quench with $\tau_Q=1$.
}
\label{fig4}
\end{figure*}
%%%%%%%%%%%%%%%%%%%%%%%%%%%%%%%%%%%%%%%%%%%%%%%%%%%%%%%
%

%%%%%%%%%%%%%%%%%%%%%%%%%%%%%%%%%%%%%%%%%%%%%%%%%%%%%%%%%%%%%%%%%%%%%%%%%%%%%%%%%%%%%%%%%%%%%%%%%%%%%%%%%%%%%%%%%%%%%%%%%%%%%%%%%%%%%%%%

\section{Impact of Environmental Noise: Noisy dynamics }
Noise is everywhere and unavoidable in any physical system. 
Specifically, when energy is moved into or out of a system in the lab, there will always be some fluctuations, ``noise", in this process. 
In addition, in the Ramsey interference scheme, the central qubit used as a noise spectrometer \cite{Bylander20,Zhang2020,Jurcevic2022,Asadian2014}. 
The noise is caused by a classical fluctuating field.
In this section, we look at how noise affects the decoherence of a qubit that's linked to a time-varying environmental spin system.
To this end, we consider an added noise to the time dependent magnetic field
\begin{equation}
h(t) = h_{0}(t) + S(t) = \frac{t}{\tau_Q} + S(t),
\end{equation}
in the ramp interval 
$ 
[t_i = h_i \tau_Q, 
~
t_f = h(t) \tau_Q],
$
where $S(t)$ represents random fluctuations with vanishing mean, $\langle S({t})\rangle=0$. 
{\color{black}This extra noisy term can resemble either a pure dephasing dynamics for the central spin.}
We assume the noise distribution is Gaussian with two-point correlations (Ornstein-Uhlenbeck process)
$$
\langle S(t)S(t')\rangle=\frac{\xi^{2}}{2\tau_n}
e^{- \frac{ |{t}-{t}'| }{ \tau_n} },
$$
 where
$\xi$ characterizes the strength of the noise and $\tau_n$ is the noise correlation time
\cite{luczka1991quantum,Budini2000,Filho2017,Kiely2021,Jafari2024a}. 
White noise is approximately equivalent to fast colored noise ($\tau_n \rightarrow 0$)  with two-point correlations 
$$
\langle S(t)S(t')\rangle=\xi^2 \delta (t-t').
$$
The mean transition probabilities {\color{black}over the whole noise distribution $S(t)$} are obtained numerically using the exact master equation~\cite{luczka1991quantum,Budini2000,Filho2017,Kiely2021,Jafari2024a} 
for the averaged density matrix
$\rho_{k}(t)$ of the noisy Hamiltonian
\bea
\bl
H^{(\xi)}_{k}(t)=H^{(0)}_{k}(t)+S(t)H^{(1)}_{k},
\el
\eea
i.e.,
%
%%%%%%%%%%%%%%%%%%%%%%%%%%%% Eq. 13   %%%%%%%%%%%%%%%%%%%%%%%%%%%%%%%%%%%
\begin{eqnarray}
\label{eq13}
\dot{\rho}_k(t)=-i\left[H^{(0)}_{k}(t),\rho_k(t)\right]-\frac{\xi^{2}}{2\tau_n}\left[H^{(1)}_{k},\int_{t_i}^{t}e^{-|t-s|/\tau_{n}}[H^{(1)}_{k},\rho_k(s)]ds\right].
\end{eqnarray}
%%%%%%%%%%%%%%%%%%%%%%%%%%%%%%%%%%%%%%%%%%%%%%%%%%%%%%%%%%%%%%%%%%%%%%
%
where  $H^{(0)}_{k}(t)$ represents the noise-free Hamiltonian and $H^{(1)}_{k} = -\sigma^{z}$ denotes the noisy component~\cite{Jafari2024a}.
%
%where $H^{(0)}_{k}(t)$ is noise-free Hamiltonian while $H^{(1)}_{k}$ expresses the ``noisy" part which is $H^{(1)}_{k}= -\sigma^{z}$ in this paper. In the quench interval $t \in\,  [t_i,t_f]$  

As described in Supplemental Material, by converting Eq.~(\ref{eq13}) into a pair of coupled differential equations, 
we numerically compute the mean values of $|u_{k}^{\pm}(t)|^2$, $|v_{k}^\pm(t)|^2$, $u_{k}^{\pm}(t)v_{k}^{\pm*}(t)$, and $u_{k}^{\pm*}(t)v_{k}^{\pm}(t)$ as ensemble averages over the noise distribution $S(t)$. It is important to note that in the limit as $\tau_n \rightarrow 0$, the above master equation simplifies to the white noise master equation
%
%%%%%%%%%%%%%%%%%%%%%%%%%%%%% Eq. 14 %%%%%%%%%%%%%%%%%%%%%%%%%%%%%%%%
{\small
\begin{equation}
\label{eq14}
\dot{\rho}_k(t) = -i\left[H^{(0)}_{k}(t), \rho_k(t)\right] - \frac{\xi^2}{2} \left[ H^{(1)}_{k}, \left[ H^{(1)}_{k}, \rho_k(t) \right] \right].
\end{equation}
}
%%%%%%%%%%%%%%%%%%%%%%%%%%%%%%%%%%%%%%%%%%%%%%%%%%%%%%%%%%%%%%

%
%%%%%%%%%%%%%%%%%%%%%%%  Fig.5   %%%%%%%%%%%%%%%%%%%%%%%
\begin{figure*}
\begin{minipage}{\linewidth}
\centerline{\includegraphics[width=0.33\linewidth]{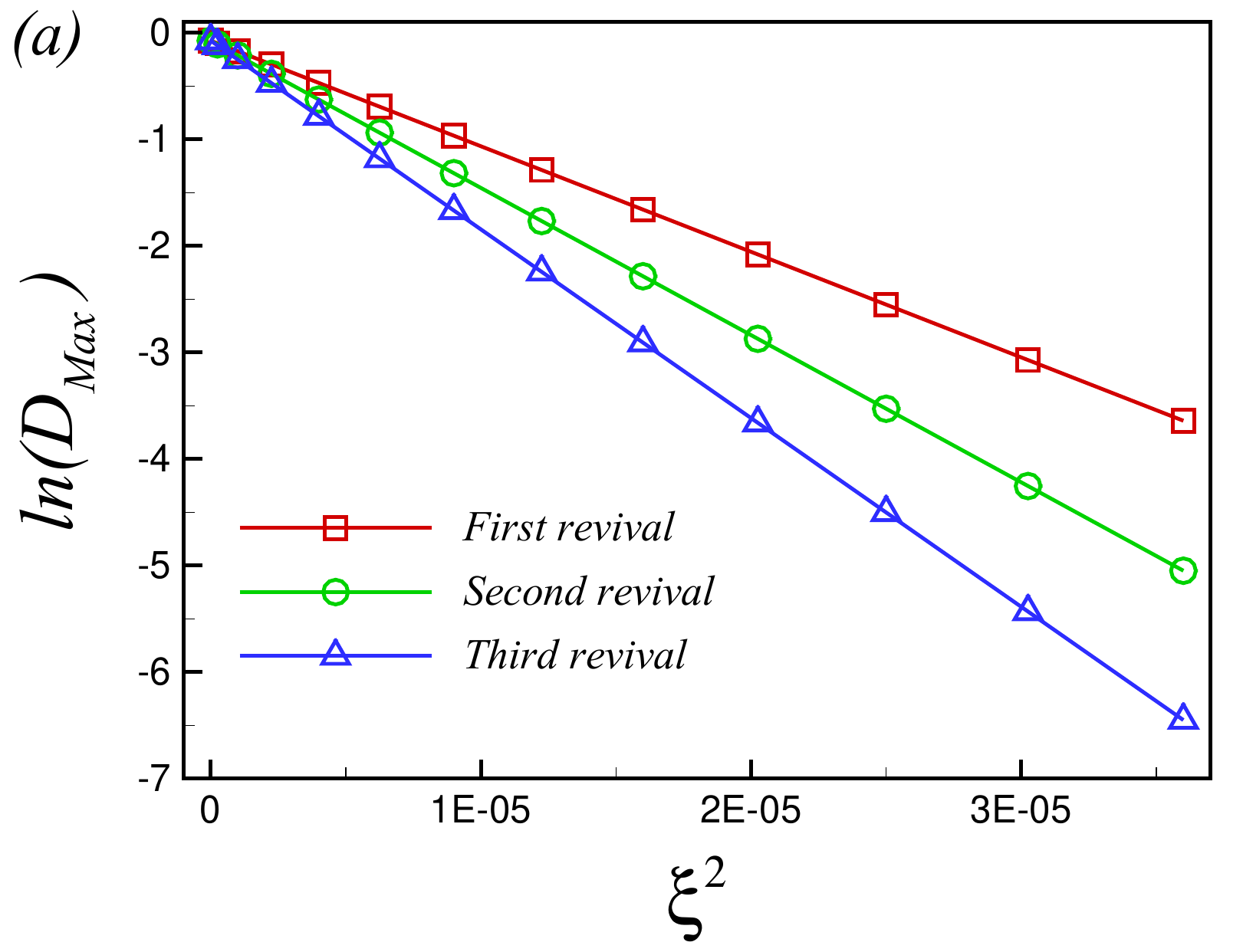}
\includegraphics[width=0.33\linewidth]{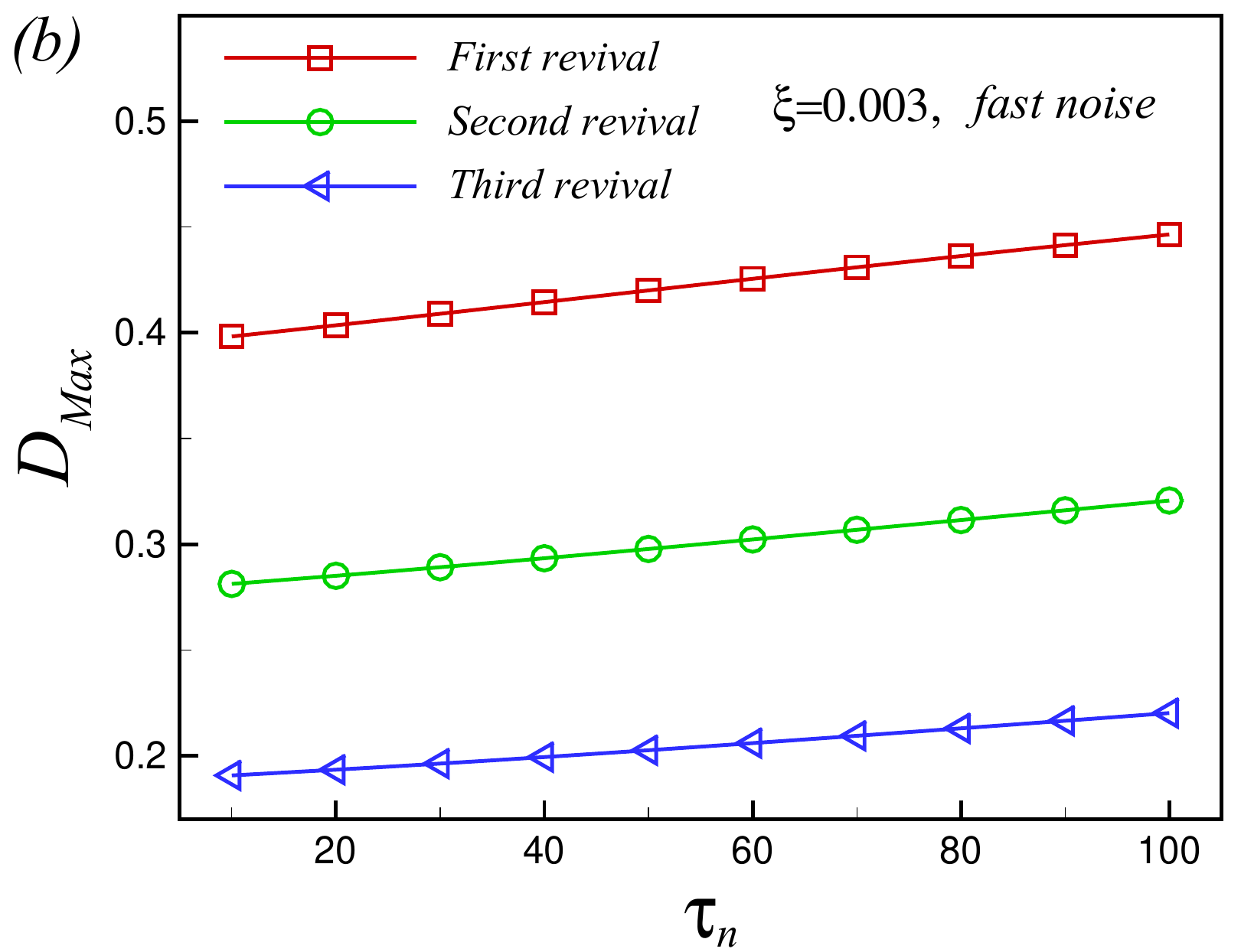}
\includegraphics[width=0.33\linewidth]{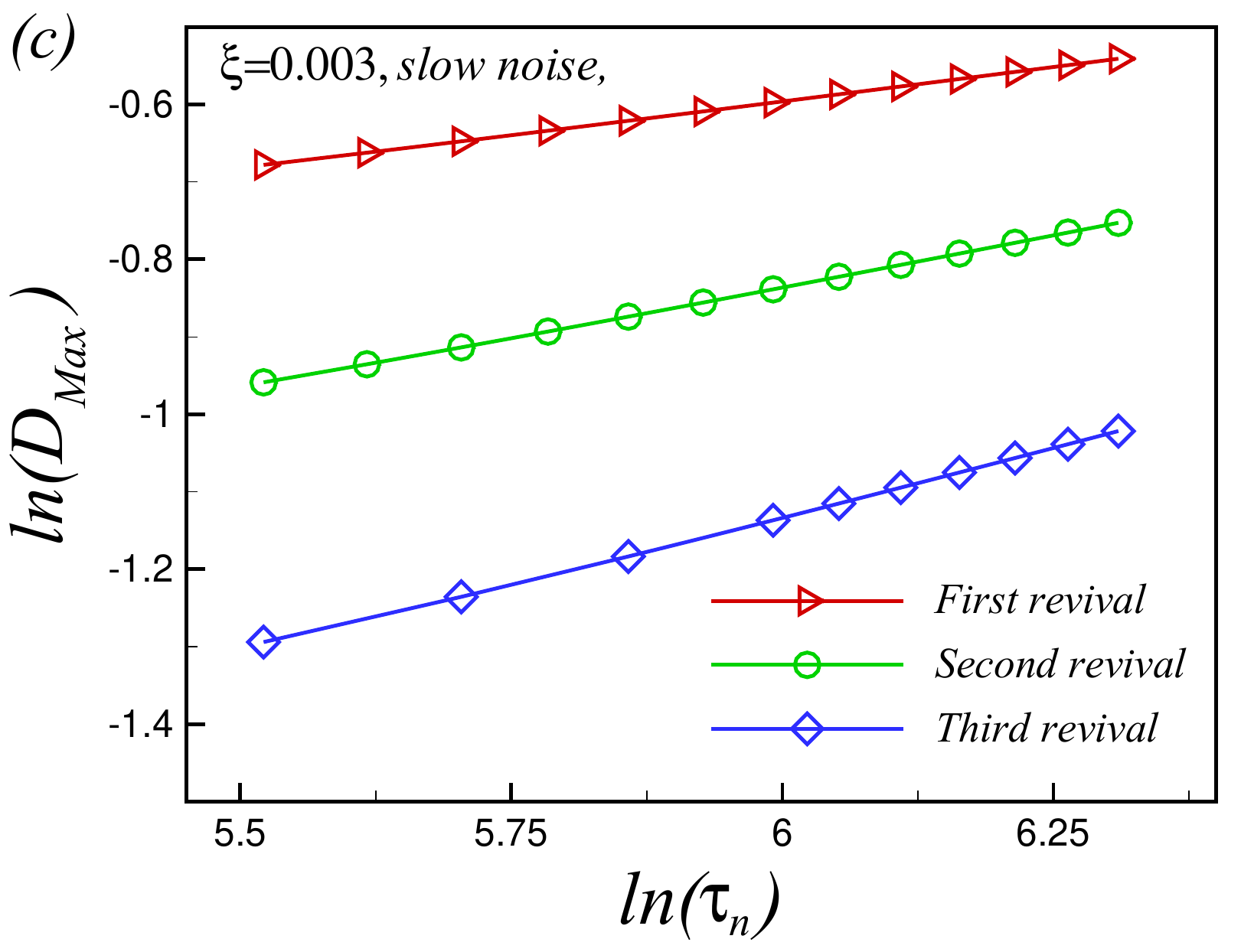}}
\centering
\end{minipage}
%===============================================================
\caption{%(Color online) 
Scaling of the maximums (revivals) of decoherence
($D_{Max}$) at the presence of the white noise and colored noise for $\tau_Q=250$. 
(a) Linear scaling of $\ln(D_{Max})$ with square of noise intensity $\xi^2$ 
corresponding to Fig.~\ref{fig2}(a).
(b) Linear scaling of revivals ($D_{Max}$) for noise intensity $\xi=0.003$ with noise 
correlation $\tau_n$ for fast noise corresponding to Fig.~\ref{fig4}(a).
(c) Scaling of $\ln(D_{Max})$ versus $\ln(\tau_n)$ for slow noise corresponding 
to Fig.~\ref{fig4}(a) for noise intensity $\xi=0.003$.}
\label{fig5}
\end{figure*}
%%%%%%%%%%%%%%%%%%%%%%%%%%%%%%%%%%%%%%%%%%%%%%%%%%%%%%%
%

%%%%%%%%%%%%%%%%%%%%%%%%%%%%%%%%%%%%%%%%%%%%%%%%%%%%%%%%%%%%%%%%%%%%%%%%%%%%%%%%%%%%%%%%%%%%%%%%%%%%%%%%%%%%%%%%%%%%%%%%%%%%

\subsection{Analysis in the presence of white noise}
Our numerical analysis begins by examining the effects of white noise on the system, as illustrated in Fig.~\ref{fig2} for ramp time scales:
(a)~$\tau_Q = 250$, 
(b)~$\tau_Q = 10$, and
(c)~$\tau_Q = 1$. 
As depicted in Fig.~\ref{fig2}(a), when the environment-qubit coupling is strong, partial revivals are still observable in the presence of the white noise, and diminish by increasing the noise intensity $\xi$. While, in the absence of noise, the revivals do not decrease by increasing
the quench time $t$, the partial revivals decay by increasing the quench time in the presence of the noise.
Decaying the revivals by increasing the quench time in the presence of the noise originates from the accumulation of noise-induced excitations during the evolution.
Notably, the period of the revivals remains consistent across both noisy and noiseless cases. In scenarios with weak coupling between the environment and the qubit, as shown in Fig.~\ref{fig2}(b), noise exacerbates the monotonic decay of decoherence.

As demonstrated in Fig.~\ref{fig2}(c),  when  the ramp time scale decreases, indicating weaker coupling between the environment and the qubit, decoherence decreases and becomes less affected by noise.
Furthermore, as shown in Fig.~\ref{fig2}(b)~and~Fig.~\ref{fig2}(c), the finite-size decoherence exhibits a sharp decay at the critical points $h_c = \pm 1$ for small ramp time scales, where the environment-qubit coupling is weak. In the case of strong coupling and large ramp time scales, the decay is observed specifically at $h_c = -1$.
While for strong environment-qubit coupling, the decoherence exhibits a maximum at $h_c = 1$, as seen in Fig.~\ref{fig2}(a).
In other words, the critical points are signaled by the finite-size decoherence.
In the absence of noise, the decoherence exhibits exponential scaling with the size of the system at the critical point, i.e., 
$$D|_{h=\pm 1}\sim e^{-N}.$$
In Fig.~\ref{fig3}, we investigated the scaling behavior of the decoherence at the critical points in the presence of noise. 
We discovered that the decoherence exhibits exponential scaling with 
$N \xi^2$, such that 
$$
D|_{h = \pm 1} \sim e^{-N \xi^2}.
$$

\subsection{Analysis in the presence of colored noise}
 
The decoherence during a quench in the presence of colored (correlated) noise is plotted in Fig.~\ref{fig4} for different values of the ramp time scale.
As seen in Fig.~\ref{fig4}(a), when the environment-qubit coupling is strong enough, partial revivals are still present in the presence of colored noise, but they decrease as the noise correlation time $\tau_n$ decreases.
This means that decoherence is less affected by correlated noise than by white noise.
Moreover, the period of the revivals remains the same in both noisy and noiseless cases.
As shown in Fig.~\ref{fig4}(b), in the case of weak coupling between the environment and the qubit, noise enhances the monotonic decay of decoherence.
As the ramp time scale get smaller, which is equivalent to weaker environment-qubit coupling, the coherency increases and affected less by noise 
as illustrated in Fig.~\ref{fig4}(c). 
Furthermore, it is clear that the critical points are signaled by the finite-size decoherence. In the presence of colored noise, we investigated the scaling behavior of the decoherence at the critical points.
We found that the decoherence exhibits exponential scaling with $N \xi^2 / \tau_n$, specifically,  $$D|_{h = \pm 1} \sim e^{-\frac{ N \xi^2 }{ \tau_n} }.$$ 
In Supplemental Material more details are given on the scaling of decoherence $D$.
\\
Fig.~\ref{fig5} illustrates the scaling of the revivals (the maximum of decoherence) in the presence of white and colored noise for $\tau_Q = 250$.
As seen in Fig.~\ref{fig5}(a), the maximum of the revivals scales exponentially with the square of the noise intensity. Additionally, the decreasing slope of the lines indicates the accumulation of noise-induced excitations during the evolution, which results in the decay of the revivals as the quench time $t$ increases. The scaling of the revivals with the noise correlation time $\tau_n$ is depicted in Fig.~\ref{fig5}(b) and Fig.~\ref{fig5}(c) for the noise intensity $\xi = 0.003$.
As observed, the maximum of the decoherence increases with the noise correlation time and scales linearly with $\tau_n$ for fast noise ($\tau_n \leq 100$), while it scales exponentially for slow noise ($\tau_n \geq 250$).
%
%%%%%%%%%%%%%%%%%%%%%%%  Fig.6   %%%%%%%%%%%%%%%%%%%%%%%
\begin{wrapfigure}{r}{0.5\textwidth}
\centering
\includegraphics[width=0.45\columnwidth]{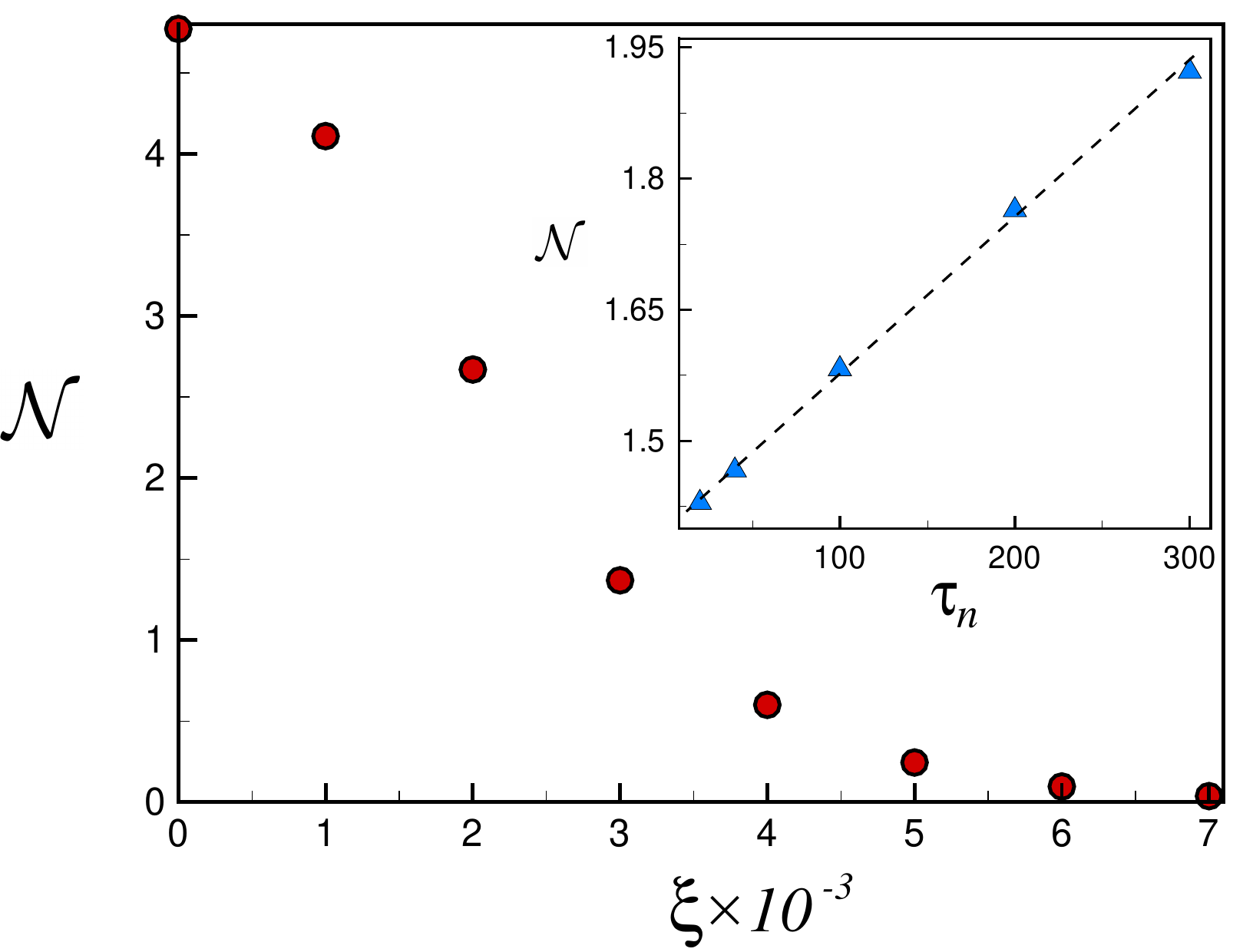}
\caption{The measure of non-Markovianity for the white noise, as a function of 
the noise strength and the inset shows the measure of non-Markovianity for the color noises
versus the noise correlation time. In the inset panel the dashed black line 
is the best linear fit to the data. Here, we have $\xi=0.003$.}
\label{fig6}
\end{wrapfigure}
%%%%%%%%%%%%%%%%%%%%%%%%%%%%%%%%%%%%%%%%%%%%%%%%%%%%%%%
% 

%%%%%%%%%%%%%%%%%%%%%%%%%%%%%%%%%%%%%%%%%%%%%%%%%%%%%%%%%%%%%%%%%%%%%%%%%%%%%%

\section{Non-Markovianity}

The revivals in decoherence of the central spin through the quench signals non-Markovianity of the dynamics.
Here, to study the features of this behavior we quantify deviation of this dynamics from a Markovian one by computing the measure proposed in Ref.~\cite{Breuer2009}.

The measure is defined based on the rate of change in the trace distance between two initially distinguishable states 
$\rho_1(0)$ and $\rho_2(0)$ for a given process that we show as $\Phi(t)$.
Indeed, for a Markovian process any two states become increasingly similar over time.
Therefore, the measure is computed based on the deviation from this property as
\begin{equation}
\mathcal{N} = \int_{>0}\hspace{-2mm} dt\ \tfrac{d}{dt}\mathcal{D}[\Phi(t)\rho_1(0),\Phi(t)\rho_2(0)],
\label{nMmeasure}
\end{equation}
where the integration is only performed for the time intervals that the integrand is positive.
Here, 
$$\mathcal{D}[\rho_1,\rho_2] = \frac{1}{2}\|\rho_1 - \rho_2 \|_1$$
 is the trace distance with $\|\cdots \|_1$ standing for the trace norm \cite{Nielsen}.
The above quantity must be maximized for different initial states.
For the pure dephasing of a central spin that we are interested in this work the superposition state already is the optimal state.

The results for the measure of non-Markovianity are shown in Fig.~\ref{fig6} for both white and colored noise cases.
In the case of white noise,
 we observe a monotonic decay in the non-Markovianity of the process as the noise strength increases.
For large enough $\xi$ values the process becomes fully Markovian.
In order to understand the role of finite noise correlation time, we compute the measure in Eq.~\eqref{nMmeasure} with different $\tau_n$ values when the noise strength is fixed to $\xi=0.003$. Interestingly, the measure exhibits a linear growth with the noise correlation time.

\section{Summary and Conclusion}
Characterizing noisy signals is crucial for understanding the dynamics of open quantum systems, enhancing our ability to control and engineer them effectively. This work investigated 
the impact of noisy environmental spin systems (ESSs) on the decoherence of the central qubit as a sensitive noise estimator compared to other methods.
We specifically focus on how noise in a time-dependent external magnetic field influences the coherence dynamics of a central spin coupled to a spin chain, particularly as the ESS is driven across its quantum critical point (QCP).

By extending the central spin model to incorporate stochastic variations in the external magnetic field, we demonstrate that noise not only amplifies decoherence resulting from the nonequilibrium critical dynamics of the environment but also profoundly affects the system's temporal evolution. Our numerical calculations reveal that decoherence exhibits exponential scaling at the critical points with both the square of the noise intensity and the noise correlation time. This contrasts with the noiseless case, where decoherence revivals occur when the chain-qubit coupling is sufficiently strong; however, these revivals diminish in the presence of noise, scaling exponentially with white noise intensity and linearly or a power law in the case of colored noise, depending on the correlation time. {\color{black} It should be emphasized that, both colored noise and white noise are classified as Gaussian noise, and a more comprehensive understanding could be achieved by exploring the effects of non-Gaussian noise on the decoherence factor.} 

Additionally, our exploration of non-Markovianity reveals that it decreases with the square of the noise intensity but increases linearly with the noise correlation time, highlighting the complex interplay between noise and memory effects in quantum systems. These findings underscore the importance of accounting for noise to model and predict quantum system behavior under realistic conditions accurately. They also offer new perspectives on the challenges and opportunities in quantum control, decoherence mitigation, and potential applications in noise spectroscopy of external signals \cite{Bylander20,Zhang2020,Jurcevic2022,Asadian2014}.

Last but not least, interference phenomena exemplified by the collapse and revival of the decoherence function serve as a direct evidence of entanglement dynamics and information flow between the central spin and the chain. But from a fundamental point of view the interference effect alone is not sufficient to rule out a classical description of the environment \cite{Leggett2002}; collapse and revival phenomena can indeed be generated by coupling the qubit to an engineered classical field \cite{Leggett1985}. Therefore, to definitively demonstrate the quantum nature of the system, a more fundamental approach is needed, similar to quantum-witness equality \cite{Ming2012} and the Leggett-Garg test \cite{Asadian2014,Leggett1985}. Therefore, the present work anticipates a systematic exploration of entanglement dynamics and quantum coherence using stronger nonclassicality criterion for distinguishing quantum behavior from classical effects in various conditions.

{\color{black} The quick advancements in the realization of analog quantum simulators indicate that we might have the opportunity to experimentally verify our predictions. Noise-averaged measurements are anticipated to be entirely feasible with current experimental techniques and are expected to provide significant insights \cite{Bylander20,Yuge2011,Fink2013,Uhrig2007,Zhang2020,Jurcevic2022}. 
Ramped magnetic quenches, conducted in the presence of amplitude-controlled noise, have already been successfully implemented with trapped ions simulating the transverse-field XY chain \cite{Ai2021}. Additionally, the foundational aspect of experimental investigation$-$detection and characterization of decoherence$-$is also established, as evidenced across various platforms for TFI-type chains with finite-range interactions, including trapped ions \cite{Jurcevic2017,Zhang2017,Nie2020}, Rydberg atoms \cite{Bernien2017}, and NV centers \cite{Chen2020}. These advancements, along with recent progress in quantum-circuit computations on NISQ devices \cite{Dborin2022}, suggest a promising avenue for exploring decoherence following noisy quenches in the nearest-neighbor interacting TFI chain discussed in this paper.}

\appendix
{\color{black}
\section{Model and exact solution\label{APC}}
The Hamiltonian of the Ising chain with periodic boundary conditions and subject to a time dependent transverse magnetic field,
%
%%%%%%%%%%%%%%%%%%%%%%%%%%%%%%%%%%%%%% Eq. Ising  %%%%%%%%%%%%%%%%%%%%%%%%%%%%%%%%%%%%%%%%%%%%
\bea
\label{eq:Ising}
H(t) =- \sum_{n=1}^{N} \big(\sigma_n^x \sigma_{n+1}^x + h(t)\sigma_n^z \big),
\eea
%%%%%%%%%%%%%%%%%%%%%%%%%%%%%%%%%%%%%%%%%%%%%%%%%%%%%%%%%%%%%%%%%%%%%%%%%%%%%%%%%%
%
where $\sigma^{x,z}_n$ are Pauli matrices acting at sites $n$ of a one-dimensional lattice. 
When the magnetic field is time-independent, $h(t)\!=\!h$, the ground state is ferromagnetic with $\langle \sigma_n^x \rangle \neq 0$ for $|h| < 1$,
otherwise paramagnetic with $\langle \sigma_n^x \rangle = 0$ , the phases being separated by quantum critical points at $h=\pm 1$ \cite{Pfeuty1970}. 

The Hamiltonian $H(t)$ in Eq. (\ref{eq:Ising}) can be mapped onto a model of spinless fermions with operators $c_n, c_n^\dagger$ using a Jordan-Wigner transformation \cite{LSM1961}
%
%%%%%%%%%%%%%%%%%%%%%%%%%%%%%%%%%%%%%%%%%  Equation.6  %%%%%%%%%%%%%%%%%%%%%%%%%%%%%%%%%%%%%%%%%%%
\bea
\label{eq:FF}
\bl
H(t)= \sum_{k>0}
[-(h(t)-\cos k)](c_{k}^{\dagger}c_{k}+c_{-k}^{\dagger}c_{-k}) +\sin(k)(c_{k}^{\dagger}c_{-k}^{\dagger}+c_{k}c_{-k}).
\el
\eea
%%%%%%%%%%%%%%%%%%%%%%%%%%%%%%%%%%%%%%%%%%%%%%%%%%%%%%%%%%%%%%%%%%%%%%%%%%%%%%%%%%%%
%

Performing a Fourier transformation, $c_n = (\mbox{e}^{i\pi/4}/\sqrt{N}) \sum_k \mbox{e}^{ikn}c_k$ (with the phase factor $\mbox{e}^{i\pi/4}$ added for convenience), 
and introducing the Nambu spinors $C_k^{\dagger} = (c_k^{\dagger} \ c_{-k})$, $H(t)$ gets expressed as a sum over decoupled mode Hamiltonians $H(t)=\sum_{k>0} C^{\dagger}_k H_{k}(t) C_k$ where, $H_{k}(t) = h_{k}(t)\sigma^z + \Delta_k\sigma^x$, with $h_{k}(t) = -( h(t)-\cos(k))$ and $\Delta_k = \sin(k)$.

For linear time dependent magnetic field, i.e., $h(t)=t/\tau_Q$ and defining  $|\varphi_{k}(t)\rangle=(v_{k}(t), u_{k}(t))^{T}$, the time dependent $v_{k}(t)$ and $u_{k}(t)$ can be calculated exactly by solving the time dependent Schr\"odinger equations \cite{Vitanov1999,Damski2011,Suzuki2016,Nag2012}
%
%%%%%%%%%%%%%%%%%%%%%%%%%%%%%%%% Eq. 8 %%%%%%%%%%%%%%%%%%%%%%%%%%%%%%%%
\begin{eqnarray}
\label{eq8}
\bl
i \frac{d}{dt} v_{k} =& -(h(t) - \cos k) v_{k} + \sin k \, u_{k},
 \\
i \frac{d}{dt} u_{k} = & (h(t)  - \cos k) u_{k} + \sin k \, v_{k},
\el
\end{eqnarray}
%%%%%%%%%%%%%%%%%%%%%%%%%%%%%%%%%%%%%%%%%%%%%%%%%%%%%%%%%%%%%%%%%%%%
%
where
$v_k(t)=U_{11}(t)v_{k}(0)+U_{12}(t)u_{k}(0)$ and $u_k(t)=U_{21}(t)v_{k}(0)+U_{22}(t)u_{k}(0)$, 
with
%
%%%%%%%%%%%%%%%%%%%%%%%%%%%%%%%%%%%%%%%%%  Eq.  %%%%%%%%%%%%%%%%%%%%%%%%%%%%%%%%%%%%%%%%%%%
{\small
\bea
\no
U_{11}(t)&=&\frac{\Gamma(1-\omega)}{\sqrt{2\pi}} 
\Big[D_{\omega-1}(-z_i)D_{\omega}(z_f)+D_{\omega-1}(z_i)D_{\omega}(-z_f)\Big]\\
\no
U_{12}(t)&=&\frac{\Gamma(1-\omega)}{\lambda\sqrt{\pi}}e^{i\pi/4}
\Big[D_{\omega}(z_i)D_{\omega}(-z_f)-D_{\omega}(-z_i)D_{\omega}(z_f)\Big]\\
\no
U_{21}(t)&=&\frac{\lambda\Gamma(1-\omega)}{2\sqrt{\pi}}e^{-i\pi/4}
\Big[D_{\omega-1}(z_i)D_{\omega-1}(-z_f)-D_{\omega-1}(-z_i)D_{\omega-1}(z_f)\Big]\\
\no
U_{22}(t)&=&\frac{\Gamma(1-\omega)}{\sqrt{2\pi}} 
\Big[D_{\omega}(-z_i)D_{\omega-1}(z_f)+D_{\omega}(z_i)D_{\omega-1}(-z_f)\Big],
\eea
}
%%%%%%%%%%%%%%%%%%%%%%%%%%%%%%%%%%%%%%%%%%%%%%%%%%%%%%%%%%%%%%%%%%%%%%%%%%%%%%%%%%%%
%
and $\Gamma(\nu)$ is Euler gamma function, $D_{\omega}(z_i)$ is the parabolic cylinder function \cite{szego1954,abramowitz1988},
$\omega=i\lambda^2/2$, $\lambda=\Delta_k\sqrt{\tau_Q}$, $z_{i}=\sqrt{2}e^{-i\pi/4}\tau_{k}(t_{i})/\sqrt{\tau_Q}$, and $\tau_k(t) = h_{k}(t)\tau_Q$.

For, $h_{k}(t) = -( h(t)\pm \delta -\cos(k))$, the wave function is defined by $|\varphi^{\pm}_{k}(t)\rangle=(v^{\pm}_{k}(t), u^{\pm}_{k}(t))^{T}$.
}

\section{Ensemble-averaged transition probabilities: Exact noise master equation\label{APA}}
As shown in Refs.~\cite{Jafari2024a,Kiely2021}, the noise-averaged density matrix $\rho_{st}(t)$ is described by the following nonperturbative exact master equation
%
%%%%%%%%%%%%%%%%%%%%%%%  Eq. Noise Master Equation %%%%%%%%%%%%%%%
\begin{eqnarray}
\label{master}
\frac{d}{dt} \rho(t) =-i [H_0(t), \rho(t)] - \frac{\xi^2}{2 \tau_n} \left[ H_1, \int_{t_i}^{t} e^{-\frac{|t-s|}{\tau_n}} [H_1(s), \rho(s)] \, ds \right].
\end{eqnarray}
%%%%%%%%%%%%%%%%%%%%%%%%%%%%%%%%%%%%%%%%%%%%%%%%
%
The Hamiltonian of the transverse field Ising model can be written as a sum of decoupled $k$-mode Hamiltonians, and consequently, the density matrix of the model has a direct product structure $\rho(t) = \otimes_{k} \rho_{k}(t)$. Therefore, the noise master equation for the ensemble-averaged density matrix $\rho_{k}(t)$ takes the form
%
%%%%%%%%%%%%%%%%%%%%%%%  Eq. k space Noise Master Equation %%%%%%%%%%%%%%%
\begin{eqnarray}
 \label{master2}
\frac{d}{dt} \rho_k(t) = -i [H_{0,k}(t), \rho_k(t)] - \frac{\xi^2}{2 \tau_n} \left[ H_1, \int_{t_i}^{t} e^{-\frac{|t-s|}{\tau_n}} [H_1(s), \rho_k(s)] \, ds \right].
\end{eqnarray}
%%%%%%%%%%%%%%%%%%%%%%%%%%%%%%%%%%%%%%%%%%%%%%%%
%
%
To solve for $\rho_k(t)$, we introduce a new operator defined by the integral in the same equation:
%%%%%%%%%%%%%%%%%%%%%%%%%%%%%%
\begin{equation}
\label{DeltaOp}
\Gamma_k(t) \equiv \int_{t_i}^{t} e^{-(t-s)/\tau_n} [H_{1}, \rho_k(s)] \, ds.
\end{equation}
%%%%%%%%%%%%%%%%%%%%%%%%%%%%%
Equation (\ref{master}) then takes the form:
%
%%%%%%%%%%%%%%%%%%%%%%%%%%%%%%%%%%%%%%%%%  Eq. S15  %%%%%%%%%%%%%%%%%%%%%%%%%%%%%%%%%%%%%%%%%%%
\begin{equation}
\label{eqAPB1}
\dot{\rho}_k(t) = -i [H^{(0)}_k(t), \rho_k(t)] - \frac{\xi^2}{2 \tau_n} [H_{1}, \Gamma_k(t)].
\end{equation}
%%%%%%%%%%%%%%%%%%%%%%%%%%%%%%%%%%%%%%%%%%%%%%%%%%%%%%%%%%%%%%%%%%%%%%%%%%%%%%%%%%%%
%
By using the Leibniz integral rule, one obtains the derivative of $\Gamma_k(t)$ with respect to time as:
%
%%%%%%%%%%%%%%%%%%%%%%%%%%%%%%%%%%%%%%%%%  Eq. S16  %%%%%%%%%%%%%%%%%%%%%%%%%%%%%%%%%%%%%%%%%%%
\begin{equation}
\label{eqAPB2}
\dot{\Gamma}_k(t) = -\frac{\Gamma_k(t)}{\tau_n} + [H_{1}, \rho_k(t)].
\end{equation}
%%%%%%%%%%%%%%%%%%%%%%%%%%%%%%%%%%%%%%%%%%%%%%%%%%%%%%%%%%%%%%%%%%%%%%%%%%%%%%%%%%%%
%
%
The elements of the ensemble-averaged density matrix $\rho_k(t)$ can now be obtained by numerically solving the coupled differential equations (\ref{eqAPB1}) and (\ref{eqAPB2}) with the initial conditions:
%%%%%%%%%%%%%
\begin{equation}
\label{initial1}
\rho_{k}(t_i) = 
\left(\begin{array}{cc}
1 & 0 \\
0 & 0 \\
\end{array}\right),
 \quad \text{and}
 \quad
\Gamma_k(t_i) = 
\left(\begin{array}{cc}
0 & 0 \\
0 & 0 \\
\end{array}\right).
\end{equation}
%%%%%%%%%%%%
In the chosen basis, the first initial condition indicates that the system is initialized in the ground state, while the second condition signifies that the initial state is noiseless.
Having obtained the ensemble-averaged density matrix 
$\rho_k(t)$ for a mode $k$ from the master equation (\ref{master}), we can obtain
the mean values of 
\begin{equation}
\bl
&
|u_{k}^{\pm}(t)|^2=
\rho_{k,(2,2)}(t), 
\\&
|v _{k}^\pm(t)|^2=
\rho_{k,(1,1)}(t), 
\\&
u_{k}^{\pm}(t)v_{k}^{\pm*}(t)
=
\rho_{k,(1,2)}(t),
\\&
u_{k}^{\pm*}(t)v_{k}^{\pm}(t)
=
\rho_{k,(2,1)}(t). 
\el
\end{equation}
For comprehensive discussions and detailed expositions on the exact noise master equations, including the formal properties of the time-evolved averaged density matrix, we direct readers to Refs.~\cite{luczka1991quantum,Kiely2021,Budini2000,Filho2017,Chenu2017}.

%%%%%%%%%%%%%%%%%%%%%%%  Fig. APB1   %%%%%%%%%%%%%%%%%%%%%%%
\begin{figure*}
\begin{minipage}{\linewidth}
\centerline{\includegraphics[width=0.33\linewidth]{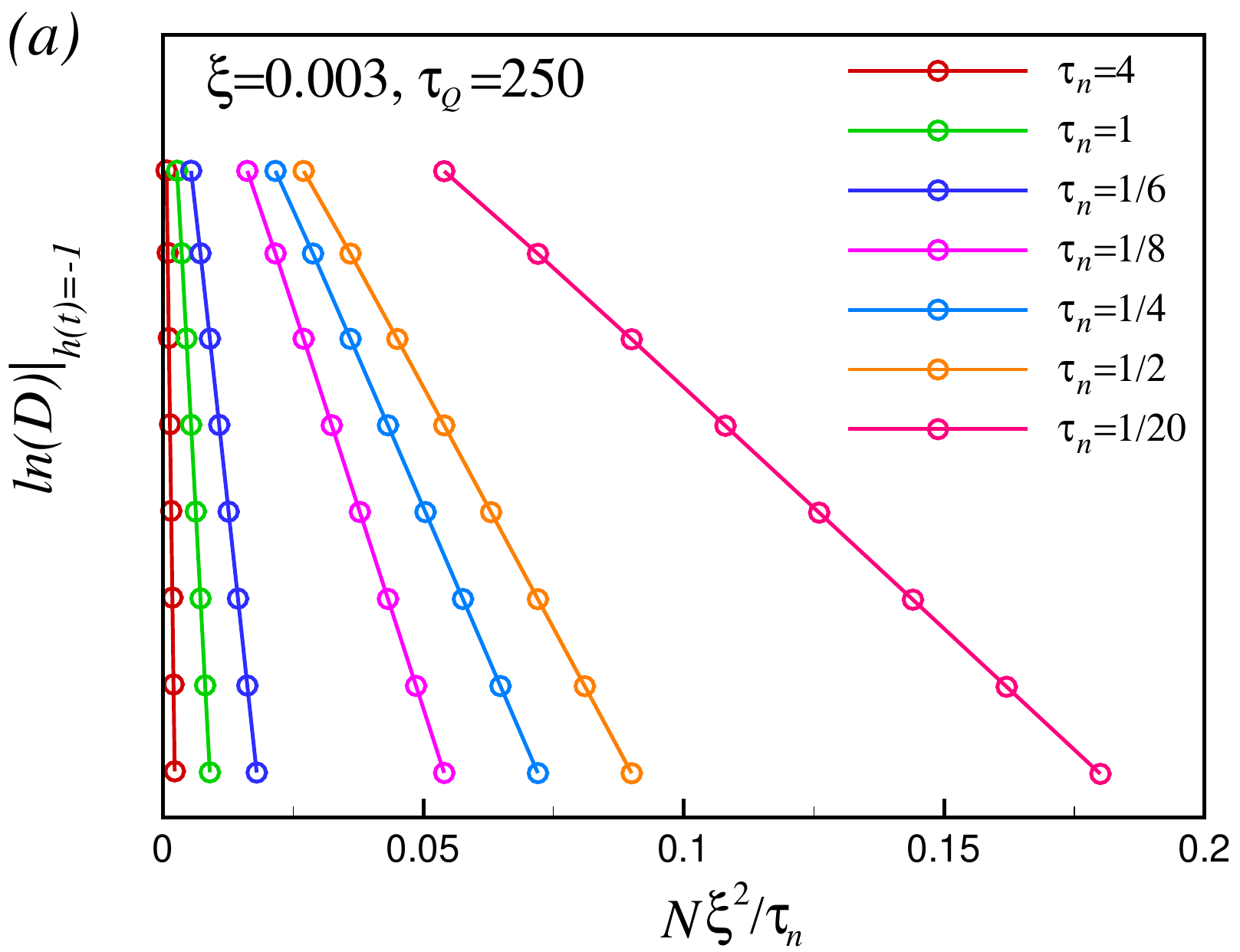}
\includegraphics[width=0.33\linewidth]{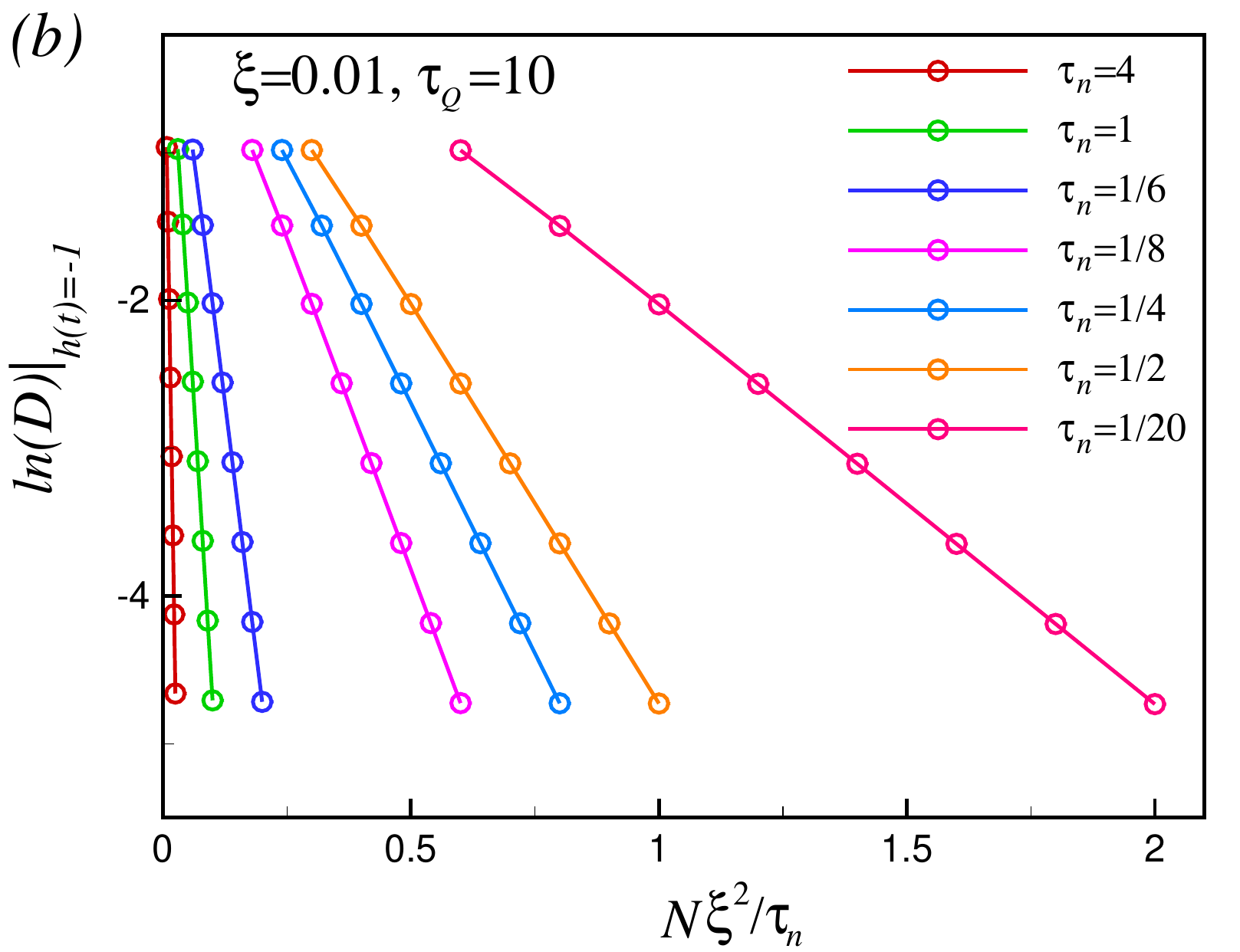}
\includegraphics[width=0.33\linewidth]{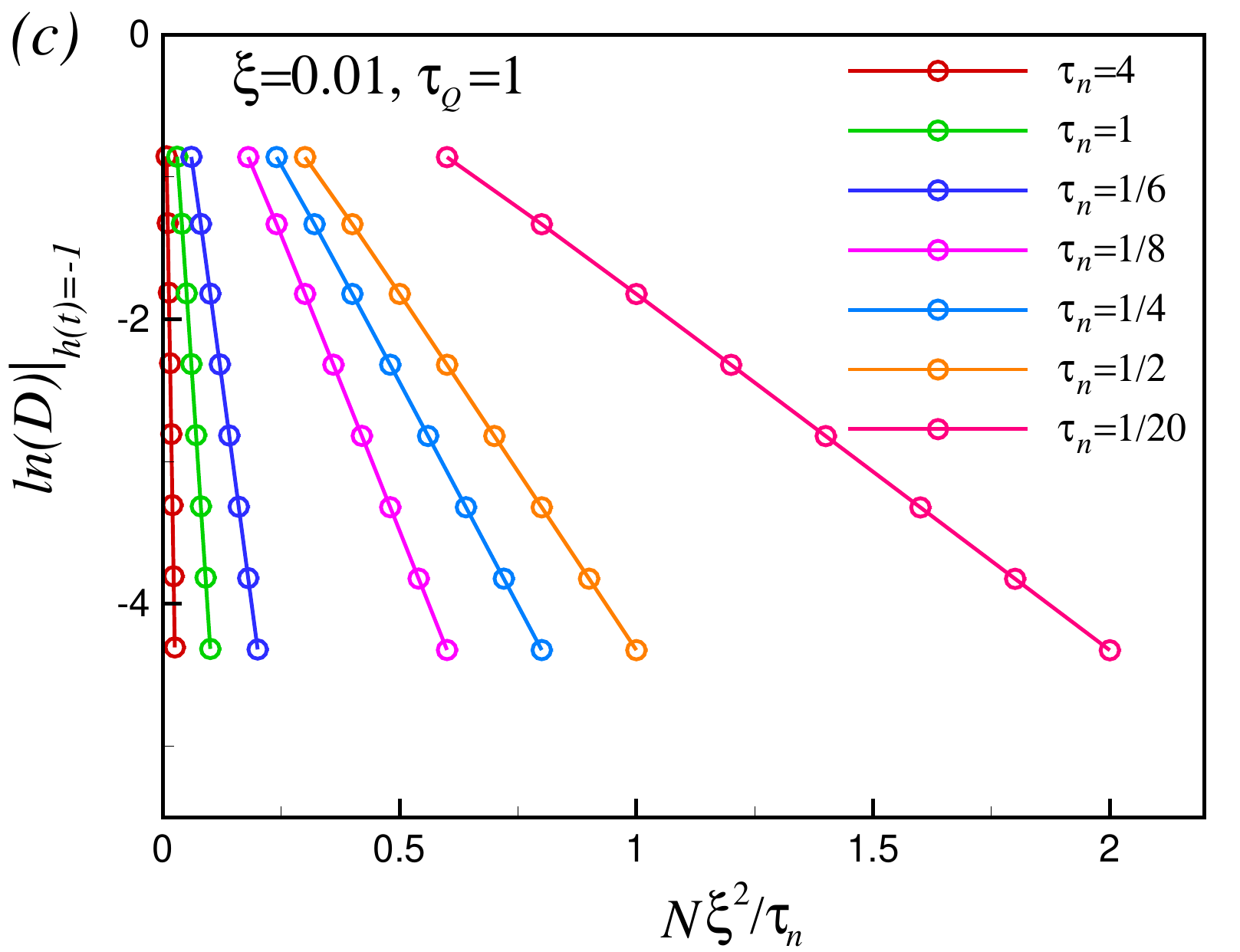}}
\centering
\end{minipage}
%===============================================================
\caption{
%(Color online) 
Scaling of decoherence
at the critical point $h_c=-1$ versus $N\xi^2/\tau_n$ for the ramp time scales:
(a) $\tau_Q=250, \xi=0.003$,
(b) $\tau_Q=10, \xi=0.01$, 
(c) $\tau_Q=1, \xi=0.01$.}
\label{figAPB1}
\end{figure*}
%%%%%%%%%%%%%%%%%%%%%%%%%%%%%%%%%%%%%%%%%%%%%%%%%%%%%%%
%
%%%%%%%%%%%%%%%%%%%%%%%%%%%%%%%%%%%%%%%%%%%%%%%%%%%%%%%%%%%%%%%%%%%%%
\section{Scaling of the decoherence factor at the critical points in the presence of colored noise\label{APB}}

As noted in the main text, decoherence scales exponentially with $N \xi^2 / \tau_n$, specifically:
$$D|_{h = \pm 1} \sim e^{-\frac{N \xi^2 }{ \tau_n} }.$$ 
This scaling behavior is depicted in Fig.~\ref{figAPB1}, which shows how decoherence at the critical point $h_c = -1$ varies with $N\xi^2/\tau_n$ for different ramp time scales.

\section*{Data availability}
All data generated or analysed during this study are included in this published article.

%%%%%%%%%%%%%%%%%%%%%%%%%%%%%%%%%%%%%%%%%%%%%%%%%%%%%%%%%%%%%%%%%%%%
%\longbibliography{REFDPD}
%\bibliography{DcoRef}

\begin{thebibliography}{}
\urlstyle{rm}
\expandafter\ifx\csname url\endcsname\relax
  \def\url#1{\texttt{#1}}\fi
\expandafter\ifx\csname urlprefix\endcsname\relax\def\urlprefix{URL }\fi
\expandafter\ifx\csname doiprefix\endcsname\relax\def\doiprefix{DOI: }\fi
\providecommand{\bibinfo}[2]{#2}
\providecommand{\eprint}[2][]{\url{#2}}

\end{thebibliography}


\begin{thebibliography}{10}
\urlstyle{rm}
\expandafter\ifx\csname url\endcsname\relax
  \def\url#1{\texttt{#1}}\fi
\expandafter\ifx\csname urlprefix\endcsname\relax\def\urlprefix{URL }\fi
\expandafter\ifx\csname doiprefix\endcsname\relax\def\doiprefix{DOI: }\fi
\providecommand{\bibinfo}[2]{#2}
\providecommand{\eprint}[2][]{\url{#2}}

\bibitem{Barenco1}
\bibinfo{author}{Barenco, A.} \& \bibinfo{author}{Ekert, A.~K.}
\newblock \bibinfo{journal}{\bibinfo{title}{Dense coding based on quantum
  entanglement}}.
\newblock {\emph{\JournalTitle{Journal of Modern Optics}}}
  \textbf{\bibinfo{volume}{42}}, \bibinfo{pages}{1253--1259},
  \doiprefix\url{10.1080/09500349514551091} (\bibinfo{year}{1995}).

\bibitem{Pereira}
\bibinfo{author}{Pereira, S.~F.}, \bibinfo{author}{Ou, Z.~Y.} \&
  \bibinfo{author}{Kimble, H.~J.}
\newblock \bibinfo{journal}{\bibinfo{title}{Quantum communication with
  correlated nonclassical states}}.
\newblock {\emph{\JournalTitle{Phys. Rev. A}}} \textbf{\bibinfo{volume}{62}},
  \bibinfo{pages}{042311}, \doiprefix\url{10.1103/PhysRevA.62.042311}
  (\bibinfo{year}{2000}).

\bibitem{Bera2018}
\bibinfo{author}{Bera, A.} \emph{et~al.}
\newblock \bibinfo{journal}{\bibinfo{title}{Quantum discord and its allies: a
  review of recent progress}}.
\newblock {\emph{\JournalTitle{Reports on Progress in Physics}}}
  \textbf{\bibinfo{volume}{81}}, \bibinfo{pages}{024001},
  \doiprefix\url{10.1088/1361-6633/aa872f} (\bibinfo{year}{2017}).

\bibitem{Rao2013}
\bibinfo{author}{Rao, K. R.~K.} \emph{et~al.}
\newblock \bibinfo{journal}{\bibinfo{title}{Multipartite quantum correlations
  reveal frustration in a quantum ising spin system}}.
\newblock {\emph{\JournalTitle{Phys. Rev. A}}} \textbf{\bibinfo{volume}{88}},
  \bibinfo{pages}{022312}, \doiprefix\url{10.1103/PhysRevA.88.022312}
  (\bibinfo{year}{2013}).

\bibitem{Barenco2}
\bibinfo{author}{Barenco, A.}
\newblock \bibinfo{journal}{\bibinfo{title}{Quantum physics and computers}}.
\newblock {\emph{\JournalTitle{Contemporary Physics}}}
  \textbf{\bibinfo{volume}{37}}, \bibinfo{pages}{375--389},
  \doiprefix\url{10.1080/00107519608217543} (\bibinfo{year}{1996}).

\bibitem{Grover}
\bibinfo{author}{Grover, L.~K.}
\newblock \bibinfo{journal}{\bibinfo{title}{Quantum mechanics helps in
  searching for a needle in a haystack}}.
\newblock {\emph{\JournalTitle{Phys. Rev. Lett.}}}
  \textbf{\bibinfo{volume}{79}}, \bibinfo{pages}{325--328},
  \doiprefix\url{10.1103/PhysRevLett.79.325} (\bibinfo{year}{1997}).

\bibitem{Jafari:2010aa}
\bibinfo{author}{Jafari, R.}
\newblock \bibinfo{journal}{\bibinfo{title}{Low-energy-state dynamics of
  entanglement for spin systems}}.
\newblock {\emph{\JournalTitle{Phys. Rev. A}}} \textbf{\bibinfo{volume}{82}},
  \bibinfo{pages}{052317}, \doiprefix\url{10.1103/PhysRevA.82.052317}
  (\bibinfo{year}{2010}).

\bibitem{Mishra2018}
\bibinfo{author}{Mishra, U.}, \bibinfo{author}{Cheraghi, H.},
  \bibinfo{author}{Mahdavifar, S.}, \bibinfo{author}{Jafari, R.} \&
  \bibinfo{author}{Akbari, A.}
\newblock \bibinfo{journal}{\bibinfo{title}{Dynamical quantum correlations
  after sudden quenches}}.
\newblock {\emph{\JournalTitle{Phys. Rev. A}}} \textbf{\bibinfo{volume}{98}},
  \bibinfo{pages}{052338}, \doiprefix\url{10.1103/PhysRevA.98.052338}
  (\bibinfo{year}{2018}).

\bibitem{Kaszlikowski2008}
\bibinfo{author}{Kaszlikowski, D.}, \bibinfo{author}{Sen(De), A.},
  \bibinfo{author}{Sen, U.}, \bibinfo{author}{Vedral, V.} \&
  \bibinfo{author}{Winter, A.}
\newblock \bibinfo{journal}{\bibinfo{title}{Quantum correlation without
  classical correlations}}.
\newblock {\emph{\JournalTitle{Phys. Rev. Lett.}}}
  \textbf{\bibinfo{volume}{101}}, \bibinfo{pages}{070502},
  \doiprefix\url{10.1103/PhysRevLett.101.070502} (\bibinfo{year}{2008}).

\bibitem{Einstein}
\bibinfo{author}{Einstein, A.}, \bibinfo{author}{Podolsky, B.} \&
  \bibinfo{author}{Rosen, N.}
\newblock \bibinfo{journal}{\bibinfo{title}{Can quantum-mechanical description
  of physical reality be considered complete?}}
\newblock {\emph{\JournalTitle{Phys. Rev.}}} \textbf{\bibinfo{volume}{47}},
  \bibinfo{pages}{777--780}, \doiprefix\url{10.1103/PhysRev.47.777}
  (\bibinfo{year}{1935}).

\bibitem{Bell}
\bibinfo{author}{Bell, J.~S.}
\newblock \bibinfo{journal}{\bibinfo{title}{On the einstein-podolsky-rosen
  paradox}}.
\newblock {\emph{\JournalTitle{Physics}}} \textbf{\bibinfo{volume}{1}},
  \bibinfo{pages}{195} (\bibinfo{year}{1964}).

\bibitem{Sadhukhan2016}
\bibinfo{author}{Sadhukhan, D.}, \bibinfo{author}{Prabhu, R.},
  \bibinfo{author}{Sen(De), A.} \& \bibinfo{author}{Sen, U.}
\newblock \bibinfo{journal}{\bibinfo{title}{Quantum correlations in quenched
  disordered spin models: Enhanced order from disorder by thermal
  fluctuations}}.
\newblock {\emph{\JournalTitle{Phys. Rev. E}}} \textbf{\bibinfo{volume}{93}},
  \bibinfo{pages}{032115}, \doiprefix\url{10.1103/PhysRevE.93.032115}
  (\bibinfo{year}{2016}).

\bibitem{Zurek1}
\bibinfo{author}{Zurek, W.~H.}
\newblock \bibinfo{journal}{\bibinfo{title}{Decoherence, einselection, and the
  quantum origins of the classical}}.
\newblock {\emph{\JournalTitle{Rev. Mod. Phys.}}}
  \textbf{\bibinfo{volume}{75}}, \bibinfo{pages}{715--775},
  \doiprefix\url{10.1103/RevModPhys.75.715} (\bibinfo{year}{2003}).

\bibitem{Kaiserbook}
\bibinfo{author}{Kaiser, R.}, \bibinfo{author}{Westbrook, C.} \&
  \bibinfo{author}{David, F.}
\newblock \emph{\bibinfo{title}{Coherent atomic matter waves-Ondes de matiere
  coherentes: 27 July-27 August 1999}}, vol.~\bibinfo{volume}{72}
  (\bibinfo{publisher}{Springer Science \& Business Media},
  \bibinfo{year}{2001}).

\bibitem{Jafari2017}
\bibinfo{author}{Jafari, R.} \& \bibinfo{author}{Johannesson, H.}
\newblock \bibinfo{journal}{\bibinfo{title}{Loschmidt echo revivals: Critical
  and noncritical}}.
\newblock {\emph{\JournalTitle{Phys. Rev. Lett.}}}
  \textbf{\bibinfo{volume}{118}}, \bibinfo{pages}{015701},
  \doiprefix\url{10.1103/PhysRevLett.118.015701} (\bibinfo{year}{2017}).

\bibitem{Chanda2016}
\bibinfo{author}{Chanda, T.} \emph{et~al.}
\newblock \bibinfo{journal}{\bibinfo{title}{Static and dynamical quantum
  correlations in phases of an alternating-field $xy$ model}}.
\newblock {\emph{\JournalTitle{Phys. Rev. A}}} \textbf{\bibinfo{volume}{94}},
  \bibinfo{pages}{042310}, \doiprefix\url{10.1103/PhysRevA.94.042310}
  (\bibinfo{year}{2016}).

\bibitem{Jafari2017b}
\bibinfo{author}{Jafari, R.} \& \bibinfo{author}{Johannesson, H.}
\newblock \bibinfo{journal}{\bibinfo{title}{Decoherence from spin environments:
  Loschmidt echo and quasiparticle excitations}}.
\newblock {\emph{\JournalTitle{Phys. Rev. B}}} \textbf{\bibinfo{volume}{96}},
  \bibinfo{pages}{224302}, \doiprefix\url{10.1103/PhysRevB.96.224302}
  (\bibinfo{year}{2017}).

\bibitem{Jafari2015}
\bibinfo{author}{Jafari, R.} \& \bibinfo{author}{Akbari, A.}
\newblock \bibinfo{journal}{\bibinfo{title}{Gapped quantum criticality gains
  long-time quantum correlations}}.
\newblock {\emph{\JournalTitle{Europhysics Letters}}}
  \textbf{\bibinfo{volume}{111}}, \bibinfo{pages}{10007},
  \doiprefix\url{10.1209/0295-5075/111/10007} (\bibinfo{year}{2015}).

\bibitem{Schliemann2002}
\bibinfo{author}{Schliemann, J.}, \bibinfo{author}{Khaetskii, A.~V.} \&
  \bibinfo{author}{Loss, D.}
\newblock \bibinfo{journal}{\bibinfo{title}{Spin decay and quantum
  parallelism}}.
\newblock {\emph{\JournalTitle{Phys. Rev. B}}} \textbf{\bibinfo{volume}{66}},
  \bibinfo{pages}{245303}, \doiprefix\url{10.1103/PhysRevB.66.245303}
  (\bibinfo{year}{2002}).

\bibitem{Cucchietti2005}
\bibinfo{author}{Cucchietti, F.~M.}, \bibinfo{author}{Paz, J.~P.} \&
  \bibinfo{author}{Zurek, W.~H.}
\newblock \bibinfo{journal}{\bibinfo{title}{Decoherence from spin
  environments}}.
\newblock {\emph{\JournalTitle{Phys. Rev. A}}} \textbf{\bibinfo{volume}{72}},
  \bibinfo{pages}{052113}, \doiprefix\url{10.1103/PhysRevA.72.052113}
  (\bibinfo{year}{2005}).

\bibitem{Korbicz2021}
\bibinfo{author}{Kici\ifmmode~\acute{n}\else \'{n}\fi{}ski, M.} \&
  \bibinfo{author}{Korbicz, J.~K.}
\newblock \bibinfo{journal}{\bibinfo{title}{Decoherence and objectivity in
  higher spin environments}}.
\newblock {\emph{\JournalTitle{Phys. Rev. A}}} \textbf{\bibinfo{volume}{104}},
  \bibinfo{pages}{042216}, \doiprefix\url{10.1103/PhysRevA.104.042216}
  (\bibinfo{year}{2021}).

\bibitem{Cucchietti2006}
\bibinfo{author}{Cucchietti, F.~M.}, \bibinfo{author}{Lewenkopf, C.~H.} \&
  \bibinfo{author}{Pastawski, H.~M.}
\newblock \bibinfo{journal}{\bibinfo{title}{Decay of the loschmidt echo in a
  time-dependent environment}}.
\newblock {\emph{\JournalTitle{Phys. Rev. E}}} \textbf{\bibinfo{volume}{74}},
  \bibinfo{pages}{026207}, \doiprefix\url{10.1103/PhysRevE.74.026207}
  (\bibinfo{year}{2006}).

\bibitem{Suzuki2016}
\bibinfo{author}{Suzuki, S.}, \bibinfo{author}{Nag, T.} \&
  \bibinfo{author}{Dutta, A.}
\newblock \bibinfo{journal}{\bibinfo{title}{Dynamics of decoherence: Universal
  scaling of the decoherence factor}}.
\newblock {\emph{\JournalTitle{Phys. Rev. A}}} \textbf{\bibinfo{volume}{93}},
  \bibinfo{pages}{012112}, \doiprefix\url{10.1103/PhysRevA.93.012112}
  (\bibinfo{year}{2016}).

\bibitem{Nag2012}
\bibinfo{author}{Nag, T.}, \bibinfo{author}{Divakaran, U.} \&
  \bibinfo{author}{Dutta, A.}
\newblock \bibinfo{journal}{\bibinfo{title}{Scaling of the decoherence factor
  of a qubit coupled to a spin chain driven across quantum critical points}}.
\newblock {\emph{\JournalTitle{Phys. Rev. B}}} \textbf{\bibinfo{volume}{86}},
  \bibinfo{pages}{020401}, \doiprefix\url{10.1103/PhysRevB.86.020401}
  (\bibinfo{year}{2012}).

\bibitem{Damski2011}
\bibinfo{author}{Damski, B.}, \bibinfo{author}{Quan, H.~T.} \&
  \bibinfo{author}{Zurek, W.~H.}
\newblock \bibinfo{journal}{\bibinfo{title}{Critical dynamics of decoherence}}.
\newblock {\emph{\JournalTitle{Phys. Rev. A}}} \textbf{\bibinfo{volume}{83}},
  \bibinfo{pages}{062104}, \doiprefix\url{10.1103/PhysRevA.83.062104}
  (\bibinfo{year}{2011}).

\bibitem{Quan2006}
\bibinfo{author}{Quan, H.~T.}, \bibinfo{author}{Song, Z.},
  \bibinfo{author}{Liu, X.~F.}, \bibinfo{author}{Zanardi, P.} \&
  \bibinfo{author}{Sun, C.~P.}
\newblock \bibinfo{journal}{\bibinfo{title}{Decay of loschmidt echo enhanced by
  quantum criticality}}.
\newblock {\emph{\JournalTitle{Phys. Rev. Lett.}}}
  \textbf{\bibinfo{volume}{96}}, \bibinfo{pages}{140604},
  \doiprefix\url{10.1103/PhysRevLett.96.140604} (\bibinfo{year}{2006}).

\bibitem{Kibble2007}
\bibinfo{author}{Kibble, T.}
\newblock \bibinfo{journal}{\bibinfo{title}{{Phase-transition dynamics in the
  lab and the universe}}}.
\newblock {\emph{\JournalTitle{Physics Today}}} \textbf{\bibinfo{volume}{60}},
  \bibinfo{pages}{47--52}, \doiprefix\url{10.1063/1.2784684}
  (\bibinfo{year}{2007}).

\bibitem{zurek1985cosmological}
\bibinfo{author}{Zurek, W.~H.}
\newblock \bibinfo{journal}{\bibinfo{title}{Cosmological experiments in
  superfluid helium?}}
\newblock {\emph{\JournalTitle{Nature}}} \textbf{\bibinfo{volume}{317}},
  \bibinfo{pages}{505--508}, \doiprefix\url{10.1038/317505a0}
  (\bibinfo{year}{1985}).

\bibitem{Uhrig2007}
\bibinfo{author}{Uhrig, G.~S.}
\newblock \bibinfo{journal}{\bibinfo{title}{Keeping a quantum bit alive by
  optimized $\ensuremath{\pi}$-pulse sequences}}.
\newblock {\emph{\JournalTitle{Phys. Rev. Lett.}}}
  \textbf{\bibinfo{volume}{98}}, \bibinfo{pages}{100504},
  \doiprefix\url{10.1103/PhysRevLett.98.100504} (\bibinfo{year}{2007}).

\bibitem{Fink2013}
\bibinfo{author}{Fink, T.} \& \bibinfo{author}{Bluhm, H.}
\newblock \bibinfo{journal}{\bibinfo{title}{Noise spectroscopy using
  correlations of single-shot qubit readout}}.
\newblock {\emph{\JournalTitle{Phys. Rev. Lett.}}}
  \textbf{\bibinfo{volume}{110}}, \bibinfo{pages}{010403},
  \doiprefix\url{10.1103/PhysRevLett.110.010403} (\bibinfo{year}{2013}).

\bibitem{Yuge2011}
\bibinfo{author}{Yuge, T.}, \bibinfo{author}{Sasaki, S.} \&
  \bibinfo{author}{Hirayama, Y.}
\newblock \bibinfo{journal}{\bibinfo{title}{Measurement of the noise spectrum
  using a multiple-pulse sequence}}.
\newblock {\emph{\JournalTitle{Phys. Rev. Lett.}}}
  \textbf{\bibinfo{volume}{107}}, \bibinfo{pages}{170504},
  \doiprefix\url{10.1103/PhysRevLett.107.170504} (\bibinfo{year}{2011}).

\bibitem{Budini2001}
\bibinfo{author}{Budini, A.~A.}
\newblock \bibinfo{journal}{\bibinfo{title}{Quantum systems subject to the
  action of classical stochastic fields}}.
\newblock {\emph{\JournalTitle{Physical Review A}}}
  \textbf{\bibinfo{volume}{64}}, \bibinfo{pages}{052110},
  \doiprefix\url{10.1103/PhysRevA.64.052110} (\bibinfo{year}{2001}).

\bibitem{Yang2017}
\bibinfo{author}{Yang, W.}, \bibinfo{author}{Ma, W.-L.} \&
  \bibinfo{author}{Liu, R.-B.}
\newblock \bibinfo{journal}{\bibinfo{title}{Quantum many-body theory for
  electron spin decoherence in nanoscale nuclear spin baths}}.
\newblock {\emph{\JournalTitle{Reports on Progress in Physics}}}
  \textbf{\bibinfo{volume}{80}}, \bibinfo{pages}{016001},
  \doiprefix\url{10.1088/0034-4885/80/1/016001} (\bibinfo{year}{2016}).

\bibitem{Montiel2013}
\bibinfo{author}{Le\'on-Montiel, R. d.~J.} \& \bibinfo{author}{Torres, J.~P.}
\newblock \bibinfo{journal}{\bibinfo{title}{Highly efficient noise-assisted
  energy transport in classical oscillator systems}}.
\newblock {\emph{\JournalTitle{Phys. Rev. Lett.}}}
  \textbf{\bibinfo{volume}{110}}, \bibinfo{pages}{218101},
  \doiprefix\url{10.1103/PhysRevLett.110.218101} (\bibinfo{year}{2013}).

\bibitem{Chenu2017}
\bibinfo{author}{Chenu, A.}, \bibinfo{author}{Beau, M.}, \bibinfo{author}{Cao,
  J.} \& \bibinfo{author}{del Campo, A.}
\newblock \bibinfo{journal}{\bibinfo{title}{Quantum simulation of generic
  many-body open system dynamics using classical noise}}.
\newblock {\emph{\JournalTitle{Phys. Rev. Lett.}}}
  \textbf{\bibinfo{volume}{118}}, \bibinfo{pages}{140403},
  \doiprefix\url{10.1103/PhysRevLett.118.140403} (\bibinfo{year}{2017}).

\bibitem{Spanner2009}
\bibinfo{author}{Spanner, M.}, \bibinfo{author}{Franco, I.} \&
  \bibinfo{author}{Brumer, P.}
\newblock \bibinfo{journal}{\bibinfo{title}{Coherent control in the classical
  limit: Symmetry breaking in an optical lattice}}.
\newblock {\emph{\JournalTitle{Phys. Rev. A}}} \textbf{\bibinfo{volume}{80}},
  \bibinfo{pages}{053402}, \doiprefix\url{10.1103/PhysRevA.80.053402}
  (\bibinfo{year}{2009}).

\bibitem{Iwakura2017}
\bibinfo{author}{Iwakura, A.}, \bibinfo{author}{Matsuzaki, Y.} \&
  \bibinfo{author}{Kondo, Y.}
\newblock \bibinfo{journal}{\bibinfo{title}{Engineered noisy environment for
  studying decoherence}}.
\newblock {\emph{\JournalTitle{Phys. Rev. A}}} \textbf{\bibinfo{volume}{96}},
  \bibinfo{pages}{032303}, \doiprefix\url{10.1103/PhysRevA.96.032303}
  (\bibinfo{year}{2017}).

\bibitem{Pichler}
\bibinfo{author}{Pichler, H.}, \bibinfo{author}{Schachenmayer, J.},
  \bibinfo{author}{Simon, J.}, \bibinfo{author}{Zoller, P.} \&
  \bibinfo{author}{Daley, A.~J.}
\newblock \bibinfo{journal}{\bibinfo{title}{Noise- and disorder-resilient
  optical lattices}}.
\newblock {\emph{\JournalTitle{Phys. Rev. A}}} \textbf{\bibinfo{volume}{86}},
  \bibinfo{pages}{051605}, \doiprefix\url{10.1103/PhysRevA.86.051605}
  (\bibinfo{year}{2012}).

\bibitem{Chen2010}
\bibinfo{author}{Chen, X.} \emph{et~al.}
\newblock \bibinfo{journal}{\bibinfo{title}{Fast optimal frictionless atom
  cooling in harmonic traps: Shortcut to adiabaticity}}.
\newblock {\emph{\JournalTitle{Phys. Rev. Lett.}}}
  \textbf{\bibinfo{volume}{104}}, \bibinfo{pages}{063002},
  \doiprefix\url{10.1103/PhysRevLett.104.063002} (\bibinfo{year}{2010}).

\bibitem{Zoller1981}
\bibinfo{author}{Zoller, P.}, \bibinfo{author}{Alber, G.} \&
  \bibinfo{author}{Salvador, R.}
\newblock \bibinfo{journal}{\bibinfo{title}{ac stark splitting in intense
  stochastic driving fields with gaussian statistics and non-lorentzian line
  shape}}.
\newblock {\emph{\JournalTitle{Phys. Rev. A}}} \textbf{\bibinfo{volume}{24}},
  \bibinfo{pages}{398--410}, \doiprefix\url{10.1103/PhysRevA.24.398}
  (\bibinfo{year}{1981}).

\bibitem{Doria2011}
\bibinfo{author}{Doria, P.}, \bibinfo{author}{Calarco, T.} \&
  \bibinfo{author}{Montangero, S.}
\newblock \bibinfo{journal}{\bibinfo{title}{Optimal control technique for
  many-body quantum dynamics}}.
\newblock {\emph{\JournalTitle{Phys. Rev. Lett.}}}
  \textbf{\bibinfo{volume}{106}}, \bibinfo{pages}{190501},
  \doiprefix\url{10.1103/PhysRevLett.106.190501} (\bibinfo{year}{2011}).

\bibitem{Marino2012}
\bibinfo{author}{Marino, J.} \& \bibinfo{author}{Silva, A.}
\newblock \bibinfo{journal}{\bibinfo{title}{Relaxation, prethermalization, and
  diffusion in a noisy quantum ising chain}}.
\newblock {\emph{\JournalTitle{Phys. Rev. B}}} \textbf{\bibinfo{volume}{86}},
  \bibinfo{pages}{060408}, \doiprefix\url{10.1103/PhysRevB.86.060408}
  (\bibinfo{year}{2012}).

\bibitem{Marino2014}
\bibinfo{author}{Marino, J.} \& \bibinfo{author}{Silva, A.}
\newblock \bibinfo{journal}{\bibinfo{title}{Nonequilibrium dynamics of a noisy
  quantum ising chain: Statistics of work and prethermalization after a sudden
  quench of the transverse field}}.
\newblock {\emph{\JournalTitle{Phys. Rev. B}}} \textbf{\bibinfo{volume}{89}},
  \bibinfo{pages}{024303}, \doiprefix\url{10.1103/PhysRevB.89.024303}
  (\bibinfo{year}{2014}).

\bibitem{Dutta2016PRL}
\bibinfo{author}{Dutta, A.}, \bibinfo{author}{Rahmani, A.} \&
  \bibinfo{author}{del Campo, A.}
\newblock \bibinfo{journal}{\bibinfo{title}{Anti-kibble-zurek behavior in
  crossing the quantum critical point of a thermally isolated system driven by
  a noisy control field}}.
\newblock {\emph{\JournalTitle{Phys. Rev. Lett.}}}
  \textbf{\bibinfo{volume}{117}}, \bibinfo{pages}{080402},
  \doiprefix\url{10.1103/PhysRevLett.117.080402} (\bibinfo{year}{2016}).

\bibitem{Bando2020}
\bibinfo{author}{Bando, Y.} \emph{et~al.}
\newblock \bibinfo{journal}{\bibinfo{title}{Probing the universality of
  topological defect formation in a quantum annealer: Kibble-zurek mechanism
  and beyond}}.
\newblock {\emph{\JournalTitle{Phys. Rev. Res.}}} \textbf{\bibinfo{volume}{2}},
  \bibinfo{pages}{033369}, \doiprefix\url{10.1103/PhysRevResearch.2.033369}
  (\bibinfo{year}{2020}).

\bibitem{Abdi2011}
\bibinfo{author}{Abdi, M.}, \bibinfo{author}{Barzanjeh, S.},
  \bibinfo{author}{Tombesi, P.} \& \bibinfo{author}{Vitali, D.}
\newblock \bibinfo{journal}{\bibinfo{title}{Effect of phase noise on the
  generation of stationary entanglement in cavity optomechanics}}.
\newblock {\emph{\JournalTitle{Phys. Rev. A}}} \textbf{\bibinfo{volume}{84}},
  \bibinfo{pages}{032325}, \doiprefix\url{10.1103/PhysRevA.84.032325}
  (\bibinfo{year}{2011}).

\bibitem{Bylander20}
\bibinfo{author}{Bylander, J.} \emph{et~al.}
\newblock \bibinfo{journal}{\bibinfo{title}{Noise spectroscopy through
  dynamical decoupling with a superconducting flux qubit}}.
\newblock {\emph{\JournalTitle{Nature Physics}}} \textbf{\bibinfo{volume}{7}},
  \bibinfo{pages}{565--570}, \doiprefix\url{10.1038/nphys1994}
  (\bibinfo{year}{2011}).

\bibitem{Zhang2020}
\bibinfo{author}{Zhang, Y.}, \bibinfo{author}{Huan, T.}, \bibinfo{author}{Zhou,
  R.-g.} \& \bibinfo{author}{Ian, H.}
\newblock \bibinfo{journal}{\bibinfo{title}{Ramsey-like spectroscopy of
  superconducting qubits with dispersive evolution}}.
\newblock {\emph{\JournalTitle{Phys. Rev. A}}} \textbf{\bibinfo{volume}{102}},
  \bibinfo{pages}{013710}, \doiprefix\url{10.1103/PhysRevA.102.013710}
  (\bibinfo{year}{2020}).

\bibitem{Jurcevic2022}
\bibinfo{author}{Jurcevic, P.} \& \bibinfo{author}{Govia, L. C.~G.}
\newblock \bibinfo{journal}{\bibinfo{title}{Effective qubit dephasing induced
  by spectator-qubit relaxation}}.
\newblock {\emph{\JournalTitle{Quantum Science and Technology}}}
  \textbf{\bibinfo{volume}{7}}, \bibinfo{pages}{045033},
  \doiprefix\url{10.1088/2058-9565/ac8cad} (\bibinfo{year}{2022}).

\bibitem{Asadian2014}
\bibinfo{author}{Asadian, A.}, \bibinfo{author}{Brukner, C.} \&
  \bibinfo{author}{Rabl, P.}
\newblock \bibinfo{journal}{\bibinfo{title}{Probing macroscopic realism via
  ramsey correlation measurements}}.
\newblock {\emph{\JournalTitle{Phys. Rev. Lett.}}}
  \textbf{\bibinfo{volume}{112}}, \bibinfo{pages}{190402},
  \doiprefix\url{10.1103/PhysRevLett.112.190402} (\bibinfo{year}{2014}).

\bibitem{Pfeuty1970}
\bibinfo{author}{Pfeuty, P.}
\newblock \bibinfo{journal}{\bibinfo{title}{The one-dimensional ising model
  with a transverse field}}.
\newblock {\emph{\JournalTitle{Annals of Physics}}}
  \textbf{\bibinfo{volume}{57}}, \bibinfo{pages}{79--90},
  \doiprefix\url{https://doi.org/10.1016/0003-4916(70)90270-8}
  (\bibinfo{year}{1970}).

\bibitem{Sun:2007aa}
\bibinfo{author}{Sun, Z.}, \bibinfo{author}{Wang, X.} \& \bibinfo{author}{Sun,
  C.~P.}
\newblock \bibinfo{journal}{\bibinfo{title}{Disentanglement in a
  quantum-critical environment}}.
\newblock {\emph{\JournalTitle{Phys. Rev. A}}} \textbf{\bibinfo{volume}{75}},
  \bibinfo{pages}{062312}, \doiprefix\url{10.1103/PhysRevA.75.062312}
  (\bibinfo{year}{2007}).

\bibitem{Sun:2010aa}
\bibinfo{author}{Sun, Z.}, \bibinfo{author}{Ma, J.}, \bibinfo{author}{Lu,
  X.-M.} \& \bibinfo{author}{Wang, X.}
\newblock \bibinfo{journal}{\bibinfo{title}{Fisher information in a
  quantum-critical environment}}.
\newblock {\emph{\JournalTitle{Phys. Rev. A}}} \textbf{\bibinfo{volume}{82}},
  \bibinfo{pages}{022306}, \doiprefix\url{10.1103/PhysRevA.82.022306}
  (\bibinfo{year}{2010}).

\bibitem{Costi:2003aa}
\bibinfo{author}{Costi, T.~A.} \& \bibinfo{author}{McKenzie, R.~H.}
\newblock \bibinfo{journal}{\bibinfo{title}{Entanglement between a qubit and
  the environment in the spin-boson model}}.
\newblock {\emph{\JournalTitle{Phys. Rev. A}}} \textbf{\bibinfo{volume}{68}},
  \bibinfo{pages}{034301}, \doiprefix\url{10.1103/PhysRevA.68.034301}
  (\bibinfo{year}{2003}).

\bibitem{lieb_two_1961}
\bibinfo{author}{Lieb, E.}, \bibinfo{author}{Schultz, T.} \&
  \bibinfo{author}{Mattis, D.}
\newblock \bibinfo{journal}{\bibinfo{title}{Two soluble models of an
  antiferromagnetic chain}}.
\newblock {\emph{\JournalTitle{Annals of Physics}}}
  \textbf{\bibinfo{volume}{16}}, \bibinfo{pages}{407--466},
  \doiprefix\url{10.1016/0003-4916(61)90115-4} (\bibinfo{year}{1961}).

\bibitem{Jafari2012}
\bibinfo{author}{Jafari, R.}
\newblock \bibinfo{journal}{\bibinfo{title}{Thermodynamic properties of the
  one-dimensional extended quantum compass model in the presence of a
  transverse field}}.
\newblock {\emph{\JournalTitle{The European Physical Journal B}}}
  \textbf{\bibinfo{volume}{85}}, \bibinfo{pages}{167},
  \doiprefix\url{10.1140/epjb/e2012-20682-5} (\bibinfo{year}{2012}).

\bibitem{Damski2005}
\bibinfo{author}{Damski, B.}
\newblock \bibinfo{journal}{\bibinfo{title}{The simplest quantum model
  supporting the kibble-zurek mechanism of topological defect production:
  Landau-zener transitions from a new perspective}}.
\newblock {\emph{\JournalTitle{Phys. Rev. Lett.}}}
  \textbf{\bibinfo{volume}{95}}, \bibinfo{pages}{035701},
  \doiprefix\url{10.1103/PhysRevLett.95.035701} (\bibinfo{year}{2005}).

\bibitem{Jafari2024a}
\bibinfo{author}{Jafari, R.}, \bibinfo{author}{Langari, A.},
  \bibinfo{author}{Eggert, S.} \& \bibinfo{author}{Johannesson, H.}
\newblock \bibinfo{journal}{\bibinfo{title}{Dynamical quantum phase transitions
  following a noisy quench}}.
\newblock {\emph{\JournalTitle{Phys. Rev. B}}} \textbf{\bibinfo{volume}{109}},
  \bibinfo{pages}{L180303}, \doiprefix\url{10.1103/PhysRevB.109.L180303}
  (\bibinfo{year}{2024}).

\bibitem{Zamani2024}
\bibinfo{author}{Zamani, S.}, \bibinfo{author}{Naji, J.},
  \bibinfo{author}{Jafari, R.} \& \bibinfo{author}{Langari, A.}
\newblock \bibinfo{journal}{\bibinfo{title}{Scaling and universality at ramped
  quench dynamical quantum phase transitions}}.
\newblock {\emph{\JournalTitle{Journal of Physics: Condensed Matter}}}
  \textbf{\bibinfo{volume}{36}}, \bibinfo{pages}{355401},
  \doiprefix\url{10.1088/1361-648X/ad4df9} (\bibinfo{year}{2024}).

\bibitem{Baghran2024}
\bibinfo{author}{Baghran, R.}, \bibinfo{author}{Jafari, R.} \&
  \bibinfo{author}{Langari, A.}
\newblock \bibinfo{journal}{\bibinfo{title}{Competition of long-range
  interactions and noise at a ramped quench dynamical quantum phase transition:
  The case of the long-range pairing kitaev chain}}.
\newblock {\emph{\JournalTitle{Phys. Rev. B}}} \textbf{\bibinfo{volume}{110}},
  \bibinfo{pages}{064302}, \doiprefix\url{10.1103/PhysRevB.110.064302}
  (\bibinfo{year}{2024}).

\bibitem{Kiely2021}
\bibinfo{author}{Kiely, A.}
\newblock \bibinfo{journal}{\bibinfo{title}{Exact classical noise master
  equations: Applications and connections}}.
\newblock {\emph{\JournalTitle{EPL}}} \textbf{\bibinfo{volume}{134}},
  \bibinfo{pages}{10001}, \doiprefix\url{10.1209/0295-5075/134/10001}
  (\bibinfo{year}{2021}).

\bibitem{Sadeghizade2025}
\bibinfo{author}{Sadeghizade, S.}, \bibinfo{author}{Jafari, R.} \&
  \bibinfo{author}{Langari, A.}
\newblock \bibinfo{journal}{\bibinfo{title}{Anti-kibble-zurek behavior in the
  quantum xy spin-$\frac{1}{2}$ chain driven by correlated noisy magnetic field
  and anisotropy}}.
\newblock {\emph{\JournalTitle{Phys. Rev. B}}} \textbf{\bibinfo{volume}{111}},
  \bibinfo{pages}{104310}, \doiprefix\url{10.1103/PhysRevB.111.104310}
  (\bibinfo{year}{2025}).

\bibitem{luczka1991quantum}
\bibinfo{author}{{\L}uczka, J.}
\newblock \bibinfo{journal}{\bibinfo{title}{Quantum open systems in a two-state
  stochastic reservoir}}.
\newblock {\emph{\JournalTitle{Czechoslov. J. Phys.}}}
  \textbf{\bibinfo{volume}{41}}, \bibinfo{pages}{289--292},
  \doiprefix\url{10.1007/BF01598768} (\bibinfo{year}{1991}).

\bibitem{Budini2000}
\bibinfo{author}{Budini, A.~A.}
\newblock \bibinfo{journal}{\bibinfo{title}{Non-markovian gaussian dissipative
  stochastic wave vector}}.
\newblock {\emph{\JournalTitle{Phys. Rev. A}}} \textbf{\bibinfo{volume}{63}},
  \bibinfo{pages}{012106}, \doiprefix\url{10.1103/PhysRevA.63.012106}
  (\bibinfo{year}{2000}).

\bibitem{Filho2017}
\bibinfo{author}{Costa-Filho, J.~I.} \emph{et~al.}
\newblock \bibinfo{journal}{\bibinfo{title}{Enabling quantum non-markovian
  dynamics by injection of classical colored noise}}.
\newblock {\emph{\JournalTitle{Phys. Rev. A}}} \textbf{\bibinfo{volume}{95}},
  \bibinfo{pages}{052126}, \doiprefix\url{10.1103/PhysRevA.95.052126}
  (\bibinfo{year}{2017}).

\bibitem{Breuer2009}
\bibinfo{author}{Breuer, H.-P.}, \bibinfo{author}{Laine, E.-M.} \&
  \bibinfo{author}{Piilo, J.}
\newblock \bibinfo{journal}{\bibinfo{title}{Measure for the degree of
  non-markovian behavior of quantum processes in open systems}}.
\newblock {\emph{\JournalTitle{Phys. Rev. Lett.}}}
  \textbf{\bibinfo{volume}{103}}, \bibinfo{pages}{210401},
  \doiprefix\url{10.1103/physrevlett.103.210401} (\bibinfo{year}{2009}).

\bibitem{Nielsen}
\bibinfo{author}{Nielsen, M.~A.} \& \bibinfo{author}{Chuang, I.~L.}
\newblock \emph{\bibinfo{title}{Quantum Computation and Quantum Information:
  10th Anniversary Edition}} (\bibinfo{publisher}{Cambridge University Press},
  \bibinfo{year}{2011}).

\bibitem{Leggett2002}
\bibinfo{author}{Leggett, A.~J.}
\newblock \bibinfo{journal}{\bibinfo{title}{Testing the limits of quantum
  mechanics: motivation, state of play, prospects}}.
\newblock {\emph{\JournalTitle{Journal of Physics: Condensed Matter}}}
  \textbf{\bibinfo{volume}{14}}, \bibinfo{pages}{R415},
  \doiprefix\url{10.1088/0953-8984/14/15/201} (\bibinfo{year}{2002}).

\bibitem{Leggett1985}
\bibinfo{author}{Leggett, A.~J.} \& \bibinfo{author}{Garg, A.}
\newblock \bibinfo{journal}{\bibinfo{title}{Quantum mechanics versus
  macroscopic realism: Is the flux there when nobody looks?}}
\newblock {\emph{\JournalTitle{Phys. Rev. Lett.}}}
  \textbf{\bibinfo{volume}{54}}, \bibinfo{pages}{857--860},
  \doiprefix\url{10.1103/PhysRevLett.54.857} (\bibinfo{year}{1985}).

\bibitem{Ming2012}
\bibinfo{author}{Li, C.-M.}, \bibinfo{author}{Lambert, N.},
  \bibinfo{author}{Chen, Y.-N.}, \bibinfo{author}{Chen, G.-Y.} \&
  \bibinfo{author}{Nori, F.}
\newblock \bibinfo{journal}{\bibinfo{title}{Witnessing quantum coherence: from
  solid-state to biological systems}}.
\newblock {\emph{\JournalTitle{Scientific reports}}}
  \textbf{\bibinfo{volume}{2}}, \bibinfo{pages}{885},
  \doiprefix\url{10.1038/srep00885} (\bibinfo{year}{2012}).

\bibitem{Ai2021}
\bibinfo{author}{Ai, M.-Z.} \emph{et~al.}
\newblock \bibinfo{journal}{\bibinfo{title}{Experimental verification of
  anti--kibble-zurek behavior in a quantum system under a noisy control
  field}}.
\newblock {\emph{\JournalTitle{Phys. Rev. A}}} \textbf{\bibinfo{volume}{103}},
  \bibinfo{pages}{012608}, \doiprefix\url{10.1103/PhysRevA.103.012608}
  (\bibinfo{year}{2021}).

\bibitem{Jurcevic2017}
\bibinfo{author}{Jurcevic~{\em et al.}, P.}
\newblock \bibinfo{journal}{\bibinfo{title}{Direct observation of dynamical
  quantum phase transitions in an interacting many-body system}}.
\newblock {\emph{\JournalTitle{Phys. Rev. Lett.}}}
  \textbf{\bibinfo{volume}{119}}, \bibinfo{pages}{080501},
  \doiprefix\url{10.1103/PhysRevLett.119.080501} (\bibinfo{year}{2017}).

\bibitem{Zhang2017}
\bibinfo{author}{Zhang, J.} \emph{et~al.}
\newblock \bibinfo{journal}{\bibinfo{title}{Observation of a many-body
  dynamical phase transition with a 53-qubit quantum simulator}}.
\newblock {\emph{\JournalTitle{Nature}}} \textbf{\bibinfo{volume}{551}},
  \bibinfo{pages}{601--604},
  \doiprefix\url{https://doi.org/10.1038/nature24654} (\bibinfo{year}{2017}).

\bibitem{Nie2020}
\bibinfo{author}{Nie, X.} \emph{et~al.}
\newblock \bibinfo{journal}{\bibinfo{title}{Experimental observation of
  equilibrium and dynamical quantum phase transitions via out-of-time-ordered
  correlators}}.
\newblock {\emph{\JournalTitle{Phys. Rev. Lett.}}}
  \textbf{\bibinfo{volume}{124}}, \bibinfo{pages}{250601},
  \doiprefix\url{10.1103/PhysRevLett.124.250601} (\bibinfo{year}{2020}).

\bibitem{Bernien2017}
\bibinfo{author}{Bernien, H.} \emph{et~al.}
\newblock \bibinfo{journal}{\bibinfo{title}{Probing many-body dynamics on a
  51-atom quantum simulator}}.
\newblock {\emph{\JournalTitle{Nature}}} \textbf{\bibinfo{volume}{551}},
  \bibinfo{pages}{579--584},
  \doiprefix\url{https://doi.org/10.1038/nature24622} (\bibinfo{year}{2017}).

\bibitem{Chen2020}
\bibinfo{author}{Chen, B.} \emph{et~al.}
\newblock \bibinfo{journal}{\bibinfo{title}{Detecting the out-of-time-order
  correlations of dynamical quantum phase transitions in a solid-state quantum
  simulator}}.
\newblock {\emph{\JournalTitle{Appl. Phys. Lett.}}}
  \textbf{\bibinfo{volume}{116}}, \bibinfo{pages}{194002},
  \doiprefix\url{10.1063/5.0004152} (\bibinfo{year}{2020}).

\bibitem{Dborin2022}
\bibinfo{journal}{\bibinfo{author}{Dborin, J.} \emph{et~al.}}
\newblock {\emph{\JournalTitle{Nat. Commun.}}} \textbf{\bibinfo{volume}{13}},
  \bibinfo{pages}{5977} (\bibinfo{year}{2022}).

\bibitem{LSM1961}
\bibinfo{author}{Lieb, E.}, \bibinfo{author}{Schultz, T.} \&
  \bibinfo{author}{Mattis, D.}
\newblock \bibinfo{journal}{\bibinfo{title}{Two soluble models of an
  antiferromagnetic chain}}.
\newblock {\emph{\JournalTitle{Ann. Phys.}}} \textbf{\bibinfo{volume}{16}},
  \bibinfo{pages}{407},
  \doiprefix\url{https://doi.org/10.1016/0003-4916(61)90115-4}
  (\bibinfo{year}{1961}).

\bibitem{Vitanov1999}
\bibinfo{author}{Vitanov, N.~V.}
\newblock \bibinfo{journal}{\bibinfo{title}{Transition times in the
  landau-zener model}}.
\newblock {\emph{\JournalTitle{Phys. Rev. A}}} \textbf{\bibinfo{volume}{59}},
  \bibinfo{pages}{988--994}, \doiprefix\url{10.1103/PhysRevA.59.988}
  (\bibinfo{year}{1999}).

\bibitem{szego1954}
\bibinfo{author}{Szeg{\"o}, G.}
\newblock \bibinfo{journal}{\bibinfo{title}{A. erd{\'e}lyi, w. magnus, f.
  oberhettinger and fg tricomi, higher transcendental functions}}.
\newblock {\emph{\JournalTitle{Bulletin of the American Mathematical Society}}}
  \textbf{\bibinfo{volume}{60}}, \bibinfo{pages}{405--408}
  (\bibinfo{year}{1954}).

\bibitem{abramowitz1988}
\bibinfo{author}{Abramowitz, M.}, \bibinfo{author}{Stegun, I.~A.} \&
  \bibinfo{author}{Romer, R.~H.}
\newblock \bibinfo{title}{Handbook of mathematical functions with formulas,
  graphs, and mathematical tables} (\bibinfo{year}{1988}).

\end{thebibliography}

%%%%%%%%%%%%%%%%%%%%%%%%%%%%%%%%%%%%%%%%%%%%%%%%%%%%%%%%%%%%%%%%%%%%

%%%%%%%%%%%%%%%%%%%%%%%%%%%%%%%%%%%%%%%%%%%%%%%%%%%%%%%%

\section*{Acknowledgement}
This work is based upon research funded by Iran National Science Foundation (INSF) under project No. 4024646.

%%%%%%%%%%%%%%%%%%%%%%%%%%%%%%%%%%%%%%%%%%%%%%%%%%%%%%%%%%%%%%%%%%%%%

\section*{Author Contributions}
All authors contribute equally to the idea generation and paper writing process.

\section*{Competing Interests}
The authors declare no competing interests.

\end{document}